\pdfoutput=1
\documentclass[opre,nonblindrev]{informs3}

\newif\ifdebug
\debugtrue

\OneAndAHalfSpacedXII

\usepackage[utf8]{inputenc}

\usepackage{endnotes}
\let\footnote=\endnote
\let\enotesize=\normalsize
\def\notesname{Endnotes}%
\def\makeenmark{\hbox to1.275em{\theenmark.\enskip\hss}}
\def\enoteformat{\rightskip0pt\leftskip0pt\parindent=1.275em
  \leavevmode\llap{\makeenmark}}

\usepackage{amssymb,amsfonts,commath}
\usepackage{dsfont,upgreek,marvosym,stmaryrd,bbold,bm}
\usepackage{fancyvrb}
\usepackage{graphicx}
\usepackage{verbatim}
\usepackage{url,hyperref}
\usepackage{float}
\usepackage{mathtools}
\usepackage{tikz}
\usetikzlibrary{arrows,matrix}
\usepackage{etoolbox}
\usepackage[perpage,para,symbol*]{footmisc}
\usepackage{multirow}
\usepackage{xifthen}
\usepackage{xspace}
\usepackage{subfig}
\usepackage[ruled]{algorithm2e}

\usepackage[noend]{algorithmic}
\usepackage{booktabs}
\usepackage{wrapfig}
\usepackage[inline]{enumitem}
\graphicspath{{./fig/}}

\SetAlFnt{\small}
\SetAlCapFnt{\small}
\SetAlCapNameFnt{\small}
\SetAlCapHSkip{0pt}
\IncMargin{-\parindent}

\DeclarePairedDelimiter{\ceil}{\lceil}{\rceil}
\DeclarePairedDelimiter{\floor}{\lfloor}{\rfloor}

\DeclarePairedDelimiter{\fracpart}{\{}{\}}

\definecolor{myblue}{HTML}{1559CF}
\definecolor{myred}{HTML}{DF1520}
\definecolor{mygreen}{HTML}{467C0D}

\ifdebug

\fi

\newcommand{\onebb}[0]{\mathds{1}}
\newcommand{\zerobb}[0]{\mathbb{0}}

\newcommand{\bigoh}[0]{\ensuremath{O}}
\newcommand{\hence}[0]{\Rightarrow}

\renewcommand{\emptyset}[0]{\varnothing}
\renewcommand{\alpha}[0]{\upalpha}
\renewcommand{\beta}[0]{\upbeta}
\renewcommand{\lambda}[0]{\uplambda}
\renewcommand{\sigma}[0]{\upsigma}

\newcommand{\const}[0]{\ensuremath{\operatorname{const}}}

\DeclareMathOperator{\E}{\ensuremath{\mathbb{E}}}

\DeclareMathOperator{\Var}{\operatorname{Var}}
\DeclareMathOperator*{\pr}{\mathbb{P}}
\input{widebar.tex}
\renewcommand{\qed}[0]{\nobreak\hfill\quad\hbox{$\blacksquare$}}  

\newcommand{\mysetminusD}{\hbox{\tikz{\draw[line width=0.6pt,line cap=round] (3pt,0) -- (0,6pt);}}}
\newcommand{\mysetminusT}{\mysetminusD}
\newcommand{\mysetminusS}{\hbox{\tikz{\draw[line width=0.45pt,line cap=round] (2pt,0) -- (0,4pt);}}}
\newcommand{\mysetminusSS}{\hbox{\tikz{\draw[line width=0.4pt,line cap=round] (1.5pt,0) -- (0,3pt);}}}
\newcommand{\setminusx}{\mathbin{\mathchoice{\mysetminusD}{\mysetminusT}{\mysetminusS}{\mysetminusSS}}}

\newcommand{\ExternalLink}{%
    \tikz[x=1.2ex, y=1.2ex, baseline=-0.05ex]{%
        \begin{scope}[x=1ex, y=1ex]
            \clip (-0.1,-0.1) 
                --++ (-0, 1.2) 
                --++ (0.6, 0) 
                --++ (0, -0.6) 
                --++ (0.6, 0) 
                --++ (0, -1);
            \path[draw, 
                line width = 0.5, 
                rounded corners=0.5] 
                (0,0) rectangle (1,1);
        \end{scope}
        \path[draw, line width = 0.5] (0.5, 0.5) 
            -- (1, 1);
        \path[draw, line width = 0.5] (0.6, 1) 
            -- (1, 1) -- (1, 0.6);
        }}

\newcommand{\vso}[0]{}
\newcommand{\vsa}[0]{\vso\vso}

\newcommand{\vsb}[0]{\vsa\vsa}

\usepackage{natbib}
 \bibpunct[, ]{(}{)}{,}{a}{}{,}%
 \def\bibfont{\small}%

\TheoremsNumberedThrough     
\EquationsNumberedThrough    

\MANUSCRIPTNO{\#}

\begin{document}

\RUNAUTHOR{Victor Amelkin and Rakesh Vohra}

\RUNTITLE{Yield Uncertainty and Formation of Supply Chain Networks}

\TITLE{Yield Uncertainty and Strategic Formation\\of Supply Chain Networks}

\ARTICLEAUTHORS{%
\AUTHOR{Victor Amelkin}
\AFF{University of Pennsylvania, Philadelphia, PA, US, \EMAIL{vctr@seas.upenn.edu}}
\AUTHOR{Rakesh Vohra}
\AFF{University of Pennsylvania, Philadelphia, PA, US, \EMAIL{rvohra@seas.upenn.edu}}
}



\MSCCLASS{90B15, 91B24, 60K10}

\ABSTRACT{%
How does supply uncertainty affect the structure of supply chain networks? To answer this question we consider a setting where retailers and suppliers must establish a costly relationship with
each other prior to engaging in trade. Suppliers, with uncertain yield, announce
wholesale prices, while retailers must decide which suppliers to
link to based on their wholesale prices. Subsequently, retailers compete with each other in Cournot
fashion to sell the acquired supply to consumers.
We find that in equilibrium 
retailers concentrate their links among too few suppliers, i.e., there is insufficient diversification of the supply base. We find that either reduction of
supply variance or increase of mean supply, increases a supplier's profit. However, these
two ways of improving service have qualitatively different effects on welfare: improvement of the
expected supply by a supplier makes everyone better off, whereas improvement of supply
variance lowers consumer surplus.
}%


\maketitle

\section{Introduction}

Semiconductors, food
processing, biopharmaceuticals, and energy are important industries that rely on suppliers subject to yield uncertainty. The degree of uncertainty can be large. \citet{bohn1999economics},
for example, suggest that disk drive manufacturer Seagate experiences production yields as low as 50\%. A popular recommendation for dealing with uncertainty on the part of suppliers is to diversify the supplier base, see \citet{chopra2006supply,cachon2008matching}. It has been widely adopted~\citep{sheffi2005resilient,sheffi2005supply}. However, signing up a new supplier  and the subsequent maintenance costs for that relationship can be costly---according to~\citet{cormican2007supplier}, it takes, on average, six months to a year to qualify
a new supplier. 

In this paper we examine the networks of buyer-supplier relationships that emerge in the presence of yield uncertainty, costly
link formation, and competition.
We consider a supply chain with many retailers (buyers) selling perfectly substitutable products
that use a common critical component. The retailers compete with each other \`{a} la Cournot,
so the market price for the retailers' output is determined by the total quantity of product present in the market. The retailers
source a common input from many unreliable suppliers, and thus face supply uncertainty. The suppliers compete on price.

Suppliers move first and set prices simultaneously. Retailers, then, simultaneously choose which
subset of suppliers to link to. Each link incurs a cost borne by the retailer. The random output of each
supplier is realized and shared equally~\citep{rong2017bullwhip,cachon1999equilibrium} between
all buyers that link to it. Random supply should be interpreted as arising from variability
in the yield of the production process. The retailers, in turn, compete downstream in Cournot 
fashion.

Retailers, in our model face a number of trade-offs. With access to more supply sources,
a retailer secures better terms of trade and is insulated against the supply uncertainty facing
any one supplier. However, given the cost of establishing a link, there is a
savings from limiting the number of suppliers. A retailer must also choose {\em which} supplier to link to, making our model
qualitatively different from such models as~\citet{mankiw1986free}, where firms decide only
upon entry and not their ``position'' within the market. On the one hand,
a supplier linked to many other retailers is unattractive because its output must be
shared with other, competing, retailers.  Yet, by coordinating on a few suppliers, retailers
can benefit from  higher downstream prices in the event that the supplier comes up
short~\citep{babich2007competition}.

We find that the resulting pure strategy Nash equilibria are inefficient in the sense of not maximizing expected welfare,
where welfare is the sum of consumer surplus, retailers' profits, and suppliers' profits. As there can be
multiple Nash equilibria, we focus on the one that minimizes the suppliers revenues.
This equilibrium employs {\em fewer} suppliers than the efficient outcome. While each retailer connects with multiple
suppliers, they concentrate their links on too few suppliers relative to the efficient number. This
tendency to agglomeration is sometimes attributed to economies of scale which are absent in our model. Rather, it is the downstream competition that drives agglomeration in our model.
Reducing the supplier base allows the retailers to earn higher prices than they otherwise would.

It is generally thought that an increase in the expected supply or a decrease in its variance should be beneficial to all. This is certainly true if a supplier increases their expected supply relative to other suppliers. Their profitability increases.
Consumer surplus also increases, and retailer profits---if we ignore indivisibilities---are unchanged.
If  a supplier reduces its 
variance relative to the other suppliers,  this strictly increases their profit, but this gain comes entirely
at the expense of consumers. Consumer surplus {\em declines} and retailer profits remain constant.
Overall welfare increases. Hence, \emph{an increase in expected supply by one supplier is unambiguously an improvement, while a reduction
in variance is not}.

In the next section of this paper we summarize the relevant prior work highlighting the main differences. The subsequent section introduces the model. The following sections provide an analysis of various parts of the model.

\section{Prior Work}

This paper occupies a position in two distinct literatures. The first is on \emph{yield uncertainty}
in production. Earlier papers focused on strategies that a single firm could employ to mitigate the effect of yield uncertainty holding competition fixed---see, for example, \citet{anupindi1993diversification}, \citet{gerchak1990yield}, and~\citet{yano1995lot}.
We, in contrast, study the interaction of competing firms.

Recently, attention has turned to the interaction between prices and yield uncertainty, with
a few representative works' being~\citet{deo2009cournot}, \citet{fang2015managing},
\citet{demirel2018strategic}, and~\citet{tang2011supplier}. In these papers the competing firms
themselves are subject to yield uncertainty, which corresponds to a single-tier supply chain, while our paper involves a two-tier supply chain. Thus these paper are
unable to say anything about the extent of supplier diversification we might observe.

Such works as
\citet{babich2007competition,ang2016disruption,bimpikis2018multisourcing,bimpikis2017supply} 
that examine multi-tier supply chains, fix the pattern of links exogenously. An exception
is~\citet{demirel2018strategic} that discusses network formation, yet, with downstream prices being fixed. Our paper has both endogenous network and price formation.

The second thread of related literature deals with \emph{network formation} between buyers and capacity-constrained sellers. In the seminal paper~\citet{kranton2001theory}, costly network formation occurs
\emph{before} prices are set. Only linked buyers and sellers can trade with each other. Once the network is formed, seller-specific prices are determined so as to clear the market. Buyers in this setting only compete for suppliers which can be interpreted as buyers choosing between differentiated sellers for personal consumption only. There is no uncertainty.
In our model, network formation occurs after prices are set and buyers (retailers) compete not just for suppliers, but also in a downstream market. Finally, we incorporate uncertainty in yield.
The same authors' related work~\citet{kranton2000networks} considers demand uncertainty,
whereas our focus is on supply uncertainty.

\section{Preliminaries}

\begin{figure}
	\FIGURE
	{\includegraphics[width=0.35\linewidth]{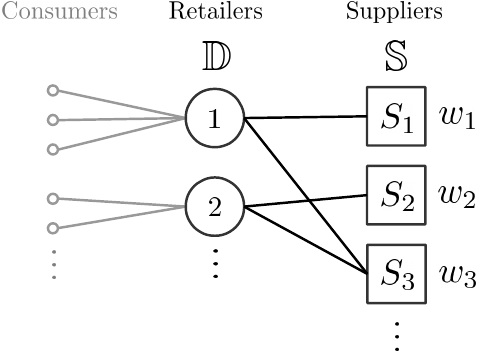}}
	{Two-tier supply chain network, with retailers $\mathbb{D}$ and suppliers
	$\mathbb{S}$. The consumer tier is implicitly present. \label{fig:supply-chain}}
	{}
\end{figure}

A \emph{supply chain} $(\mathbb{D}, \mathbb{S}, g)$, illustrated in Fig.~\ref{fig:supply-chain},
consists of $n$ retailers $\mathbb{D}$,
$m$ suppliers $\mathbb{S}$, and links $g \in \times_1^n~(2^\mathbb{S})$ between them.
Retailers and suppliers are strategic, while consumers are price-takers.
Suppliers manufacture the
product at zero marginal cost and sell it to retailers. Each supplier is
free to set any price. Based on the prices set, retailers
choose which suppliers to deal with and purchase all their output
at the price set. This is an example of a price only
contract~\citep{cachon2001contracting}. The retailers in turn sell the output to consumers
at a price determined in Cournot fashion.

The supply $S_j$ of supplier $j$ is a random variable, with mean $\E[S_j] = \mu_j$ and
variance $\Var[S_j] = \sigma^2_j$, whose value lies in $[0, S^{max}]$, $S^{max} > 0$. If the realized supply is $S$, the realized yield will be $\tfrac{S}{S^{max}}$. Unlike in~\citet{deo2009cournot} and~\citet{fang2015managing}, where $S^{\max}$ is a choice, in
our model this is fixed exogenously. One can think of $S^{max}$ as a capacity choice that over
a short time horizon is inflexible. In our model suppliers only choose a price for their realized
output.

With the exception of Sec.~\ref{sec:prices-with-varying-sigma} and~\ref{sec:prices-with-varying-mu}---we assume identically distributed supplies, with mean
$\E[S_j] = \mu$ and variance $\Var[S_j] = \sigma^2$. Supplies $S_j$ of different suppliers
are assumed to be independent. 
The maximal total amount of supply the suppliers can produce is $\Delta = m S^{max}$.

The inverse demand curve in the market downstream of the retailers is $p = \Delta - q$, where
$q$ is the total output of all the retailers, and $p$ is the market price.

Supplier $j$ sets price of $w_j$ per unit of product, and the price can vary across suppliers.
Each retailer $i$ chooses links $g_i \subseteq \mathbb{S}$ to suppliers---incurring a constant
cost of $c$ per link---thereby, forming
the supply chain network $g = (g_1, \dots, g_n)$. In that network, suppliers with at least
one link are termed \emph{active} and are denoted by $\mathbb{S}_+(g) \subseteq \mathbb{S}$,
while the suppliers having no links are termed \emph{vacant} and denoted by
$\mathbb{S}_0(g) = \mathbb{S} \setminusx \mathbb{S}_+(g)$.

 $\mathbb{Z}_+$ denotes the set of non-negative
integers, $\floor{x} \in \mathbb{Z}$ is the integer closest to and no larger than $x$ (floor),
$\fracpart{x} = x - \floor{x}$ is the fractional part of $x$, $N(k) \subseteq \{1, 2, \dots\}$
(or $N(k, g)$) is the neighborhood of node $k$ (in network $g$), and $d(k) = |N(k)|$ (or
$d(k, g)$) is the degree
of node $k$ (in network $g$). We summarize our notation in Table~\ref{tbl:notation-summary}.
\begin{table}
	\TABLE
	{Notation summary. \label{tbl:notation-summary}}
	{
    \small
    \begin{tabular}{|c|l|}
        \hline
        $\floor{x}$ & integer closest to and no larger than $x$ (floor)\\ \hline
        $\fracpart{x}$ & $x - \floor{x}$\\ \hline
        $\mathbb{D} = \{1, \dots, n\}$ & retailers, $n \gg 1$\\ \hline
        $\mathbb{S} = \{1, \dots, m\}$ & suppliers, $m \gg 1$\\ \hline
        $\mathbb{S}_+(g) \subseteq \mathbb{S}$ & active (linked) suppliers in network $g$\\ \hline
        $\mathbb{S}_0(g)$ & $\mathbb{S} \setminusx \mathbb{S}_+(g)$ -- vacant suppliers\\ \hline
        $K = K(g)$ & $|\mathbb{S}_+(g)|$ -- number of active suppliers\\ \hline
        $S_j$ & random supply of supplier $j$\\ \hline
        $\E[S_j] = \mu_j$ & expected supply of supplier $j$\\ \hline
        $\Var[S_j] = \sigma^2_j$ & supply variance of supplier $j$\\ \hline
        $\Delta$ & maximal total supply\\ \hline
        $c$ & cost of linking to a supplier\\ \hline
        $w_j$ & price supplier $j$ charges per unit of supply\\ \hline
        $g_i \subseteq \mathbb{S}$ & pure strategy of retailer $i$\\ \hline
        $g_{-i}$ & $(g_1, \dots, g_{i-1}, g_{i+1}, \dots, g_n)$\\ \hline
        $g = (g_i, g_{-i})$ & pure strategy profile / network\\ \hline
        $N(k) \subseteq \{1, 2, \dots\}$ & neighborhood of node $k$\\ \hline
        $N(k, g)$ & neighborhood of node $k$ in network $g$\\ \hline
        $d(k) = |N(k)|$ & degree of node $k$\\ \hline
    \end{tabular}
    }
    {}
\end{table}

\section{Network Formation with Fixed Prices}

In this section, we will outline the model of supply chain network formation with identical---w.r.t. supply
distributions and linking costs---suppliers, where supplier wholesale prices $(w_1, \dots, w_m)$ are assumed to be
fixed and potentially different from each other. In the next section, we extend this model to the case
where suppliers strategically select prices.

\subsection{Network Formation Game}

Given wholesale prices, $(w_1, \dots, w_m)$, we write down the payoff of retailer $i$.
\begin{definition}[Payoff of a Retailer]
At supply prices $w = (w_1, \dots, w_m)$ per unit of supply, the expected payoff of retailer $i \in \mathbb{D}$ is:
    \begin{align}
        u_i(g, w) &= \sum\limits_{j \in N(i)}{
            \Big(
                \Big(
                    \Delta - \sum_{k \in \mathbb{S}_+(g)}{S_k} - w_j
                \Big)
                \frac{S_j}{d(j)} - c
            \Big)
        },
    \label{eq:utility-of-retailer}
    \end{align}
    where $S_j \sim \operatorname{Dist}([0, S^{max}])$.
	\label{def:retailer-utility}
\end{definition}

In Definition~\ref{def:retailer-utility}, prices $w$ are announced by suppliers in advance;
additionally, variable $g$ in the expressions of type $N(i, g)$ and $d(j, g)$ is omitted for readability.

The rationale for the retailer payoff function~\eqref{eq:utility-of-retailer} is as follows.
If retailer $i$ is linked to supplier $j$, then it receives $S_j / d(j)$ amount of supply
from $j$---similarly to each out of $d(j)$ retailers linked to $j$. This supply distribution
scheme assumes that, regardless of the number of links connected to a supplier, all its supply
is realized, and that suppliers are non-discriminating in that
each supplier distributes its entire supply uniformly across the retailers linked to it.
See~\citet{rong2017bullwhip} and~\citet{cachon1999equilibrium} for a justification.

The $S_j / d(j)$ units of supply from supplier $j$ will earn retailer $i$ a marginal profit of
$(\Delta - \sum_{k \in \mathbb{S}_+(g)}{S_k} - w_j)$ per unit---the retailer purchases product
upstream from supplier $j$ at unit price $w_j$, and sells it downstream at the market price of
$(\Delta - \sum_{k \in \mathbb{S}_+(g)}{S_k})$ per unit.
Additionally, a retailer
incurs a constant cost $c$ for maintaining a link to supplier $j$.

Notice that the only way different summands in~\eqref{eq:utility-of-retailer}---corresponding
the marginal payoffs of linking to the corresponding suppliers---can affect each other is via potentially
altering the number of active suppliers $K(g) = |\mathbb{S}_+(g)| = |\{ j \in \mathbb{S} \mid d(j, g) > 0\}|$.

\begin{lemma}[Expected Payoff of a Retailer {\hyperlink{thm:expected-retailer-utility-proof}{[Proof \ExternalLink]}}]
	The expected payoff of retailer $i \in \mathbb{D}$ is
	\begin{align}
		\E[u_i(g, w)] = \sum\limits_{j \in N(i)}{\Big(
			\frac{v(K) - \mu w_j}{d(j)} - c
		\Big)},
		\label{eq:expected-utility-retailer}
	\end{align}
	where $K = K(g) = |\mathbb{S}_+(g)| = |\{ j \in \mathbb{S} \mid d(j, g) > 0\}|$
	is the number of active (linked) suppliers, and
	\begin{align}
		v(K) = \mu(\Delta - \mu K) - \sigma^2
	\end{align}
	is the ``value'' of a supplier.
	\label{thm:expected-retailer-utility}
\end{lemma}

In what follows, we define a one-shot network formation game with given wholesale prices $(w_1, \dots, w_m)$.
\begin{definition}[Network Formation Game with fixed Wholesale Prices]
	This is a one-shot
	game, where wholesale prices $(w_1, \dots, w_m)$  are assumed fixed, and retailers
	strategically form links to suppliers, maximizing expected payoff  $\E[u_i(g, w)]$ over $g_i$.
	\label{def:network-formation-game}
\end{definition}

In the above defined game, we are interested in pure strategy Nash equilibria, defined
in a standard manner as follows.

\begin{definition}[Pure Strategy Nash Equilibrium Networks]
	A network $g^*$ is said to be a \emph{pure strategy Nash equilibrium} of the  network formation game with fixed wholesale prices if for all retailers $i \in \mathbb{D}$
	and any linking pattern $g_i$ being a unilateral deviation from $g^*$, the following holds
	$$
		\E[u_i(g_i, g_{-i}^*)] \leq \E[u_i(g_i^*, g_{-i}^*)].
	$$
	\vsb
	\label{def:nash-equilibrium-network}
\end{definition}

First, notice that the best-case marginal payoff from linking to supplier $j$---deduced from
equation~\eqref{eq:expected-utility-retailer}---corresponds to
the case where this is the only active supplier in the network (so $K = 1$), and the link being
created is the only link present in the network (so $d(j) = 1$). The corresponding marginal
payoff of a retailer is $v(1) - \mu w_j - c$.
It is reasonable to assume that for every supplier, this best-case marginal payoff is non-negative,
or, alternatively, every supplier has a chance of being linked to. In order to assure that the above
expression is non-negative, we make the following assumption about the costs involved in a
retailer's payoff.

\begin{assumption}[Upper-Bounded Wholesale Prices and Linking Costs]
	We assume that the suppliers' prices are reasonably small, so that every supplier can potentially
	be linked to in the best case:
	\begin{align*}
		w_j \leq \frac{v(1) - c}{\mu} = \Delta - \mu - \frac{\sigma^2 + c}{\mu}.
	\end{align*}
	Furthermore, to make sure that the upper bound in the expression above is non-negative
	(as prices $w_j$ cannot be negative), we assume that the linking cost $c$ is also bounded
	\begin{align*}
		0 < c \leq v(1) = \mu (\Delta - \mu) - \sigma^2.
	\end{align*}
	\label{asm:bounded-prices-and-linking-costs}
	\vsb
\end{assumption}

Finally, we will make another assumption about the size of our system.

\begin{assumption}[Large Supply Chain]
	We will assume that a supply chain is large in that both the number $n$ of retailers
	and the number $m$ of suppliers are sufficiently large (yet, finite).
	\label{asm:large-supply-chain}
\end{assumption}

Assumption~\ref{asm:large-supply-chain} will allow us to clearly see the effect of supply
distribution parameters as well as costs upon the formed prices and networks rather than
the effect of those mixed together with the effect of agent scarcity.

\subsection{Nash Equilibrium Analysis}
Algorithm~\ref{alg:greedy-equil-gen} below greedily constructs an equilibrium,
which is proven in the following Lemma~\ref{thm:greedy-equil-constr}.

\newcommand{\asep}[0]{\vspace{0.04in}}
\newcommand{\CONTINUE}[0]{\mbox{\bf continue\ }}
\newcommand{\algorithmicbreak}{\textbf{break}}
\newcommand{\BREAK}{\STATE \algorithmicbreak}
\begin{algorithm}[ht]
    \caption{
        Greedy construction of a pure strategy Nash equilibrium.
        Below,
        $K = K(g^*) = |\mathbb{S}_+(g^*)|$ is the number of
        active suppliers in $g^*$, and $v(K) = \mu (\Delta - \mu K) - \sigma^2$. 
    }
    \begin{algorithmic}[1]
		\openup -0.8em
        \renewcommand{\algorithmicrequire}{\textbf{Input:}}
        \renewcommand{\algorithmicensure}{\textbf{Output:}}

        \asep
        \REQUIRE $\Delta$, $\mu$, $\sigma$, $c$, $w_1, \dots, w_m$
        
        \asep         
        \ENSURE $g^*$
        
        \asep
        \STATE $g^* \leftarrow \emptyset$, $i \leftarrow 1$
        
        \asep
        \FOR{$j \in \langle 1, \dots, m \rangle
            \text{ ordered $\nearrow$ by } w_j$}
            
            \asep
            \IF{$v(K(g^*) + 1) - \mu w_j - c < 0$}
            
				\asep
				\BREAK
                
            \ENDIF
            
				\asep
                \STATE $g^* \leftarrow g^* + (i, j)$
                
                \asep
                \STATE $i \leftarrow i + 1$
            
        \ENDFOR
        
        \asep
        \FOR{$j \in \mathbb{S}_+(g^*)$}
        
            \asep
            \WHILE{ $(v(K) - \mu w_j) / (d(j, g^*) + 1) - c \geq 0$ }
                
                \asep
                \STATE $g^* \leftarrow g^* + (i, j)$
                
                \asep
                \STATE $i \leftarrow i + 1$
            
            \ENDWHILE
        
        \ENDFOR
        
        \vspace{-0.06in}
        \RETURN $g^*$
        \openup 0.8em
     \end{algorithmic}
\label{alg:greedy-equil-gen}
\end{algorithm}
\vso

\begin{lemma}[Greedy Construction of Equilibrium]
	For a supply chain network formation game with fixed wholesale prices $w$ and
    a sufficiently large number $n$ of retailers, Algorithm~\ref{alg:greedy-equil-gen}
    always terminates and outputs a pure strategy Nash equilibrium of the game.
    \label{thm:greedy-equil-constr}
\end{lemma}
\proof{Proof of Lemma~\ref{thm:greedy-equil-constr}:}
	Algorithm~\ref{alg:greedy-equil-gen} consists of two parts corresponding to
	two for-loops: in the first part, it activates as many suppliers as possible, creating
	links from different retailers; in the second part, all these active suppliers
    receive extra links until linking to them stops being profitable.
    
    Let us, first, notice that the algorithm always terminates in $\bigoh(n m)$
    steps. The first for-loop executes no more than $m$ times. The while-loop
    (lines 8-10) executes at most $n$ times, as at every iteration the chosen
    supplier $j$ may be linked to some retailer $i$, and, eventually, 
    linking to $j$ will get too expensive (due to the fact that $v(K)$ is strictly monotonically
    decreasing). Notice that, due to the assumption about
    the number $n$ of suppliers being sufficiently large, we are guaranteed to never
    run out of vacant retailers to link to suppliers. Thus, the second for-loop
    (lines 7-10) executes at most $n m$ times. Hence, the algorithm terminates in
    at most $n m$ steps.
    
    Now, we need to show that the output $g^*$ is a Nash equilibrium. In $g^*$, only
    the following types of unilateral deviations are possible:
    (a) A retailer adds new links.
    (b) A retailer having a link drops it.
    (c) A retailer having a link drops it and adds new links.
	Deviations of type~(a) are impossible (cannot result in a higher
	expected payoff), as two for-loops ensure that creation of extra links cannot have a
	non-negative marginal payoff. Deviations of type~(b) are also
	impossible, as each retailer can at most drop a single link, and that link---due
	to the if-statement inside the first for-loop---has a non-negative marginal payoff.
	Finally, deviations of type~(c) are also impossible, as, having
	dropped the single link, a retailer can at best re-link to the same supplier,
	as all the other suppliers---both active and vacant---are ``saturated'' in that linking
	to them provides a negative marginal payoff increase---vacant suppliers are such due
	to the first for-loop, and the active suppliers are such due to the second for-loop.
	Thus, $g^*$ is a pure strategy Nash equilibrium of the game.
	\qed
\endproof

The following is now immediate.
\begin{theorem}[Equilibrium Existence]
	For a supply chain network formation game with fixed $w$ and
    a sufficiently large number $n$ of retailers, pure strategy Nash equilibrium
    always exists.
	\label{thm:network-formation-equilibrium-existence}
\end{theorem}

Algorithm~\ref{alg:greedy-equil-gen} and the accompanying Lemma~\ref{thm:greedy-equil-constr},
 also provide us with information about
the active suppliers at \emph{any} equilibrium.
\begin{theorem}[Active Suppliers at Equilibrium]
	In a supply chain network formation game with fixed wholesale prices $w$ and
	a sufficiently large number $n$ of retailers, let  $K^*$ be
	the number of active suppliers in a pure strategy Nash equilibrium
	constructed by Algorithm~\ref{thm:greedy-equil-constr}. Then, in \emph{any}
	pure strategy Nash equilibrium $g^*$ of that game,  the number of active
	suppliers is $K^*$ if $v(K^*) - \mu w_{K^*} - c > 0$, and either $K^*$ or
	$(K^* - 1)$ otherwise, where, as before, prices $w_j$ are listed in
	an ascending order.
	\label{thm:active-sups-at-equilibrium}
\end{theorem}
\proof{Proof of Theorem~\ref{thm:active-sups-at-equilibrium}:}
	First, notice that the number $K$ of active suppliers at any equilibrium $g^*$
	cannot be greater than $K^*$. If that was the case in $g$, then the marginal benefits
	of linking to the cheapest $K^*$ active suppliers in $g$ would be no greater than
	those of linking to the cheapest $K^*$ active suppliers in $g^*$, and---due to lines
	(3-4) of Algorithm~\ref{alg:greedy-equil-gen} and $v(K)$'s being strictly monotonically
	decreasing in $K$---linking to the remaining $(K - K^*)$ suppliers would have a
	negative marginal benefit.
	
	Secondly, the number of active suppliers $K$ at any equilibrium also cannot be
	lower than $K^*$ when $v(K^*) - \mu w_{K^*} - c > 0$, and lower than
	$(K^* - 1)$ otherwise, since, if that was the case, one of the cheapest $K^*$ suppliers
	would have been vacant, and there would be a retailer having no links
	(as $n$ is sufficiently large, such a retailer always exists) that would be willing to link
	to one of these still vacant cheapest suppliers (due to the first for-loop of
	Algorithm~\ref{alg:greedy-equil-gen}).
		
	Thus, the number of active suppliers at any equilibrium should be $K^*$ or
	$(K^* - 1)$.
	\qed
\endproof

We now characterize pure strategy Nash equilibria of the game with fixed $w$.

\begin{theorem}[Nash Equilibrium Network Characterization {\hyperlink{thm:network-formation-nash-equilibrium-char-proof}{[Proof \ExternalLink]}}]
		In a supply chain network formation game with fixed $w_1 \leq \dots \leq w_m$, 
		and sufficiently many retailers and suppliers, a pure
		strategy Nash equilibrium will have $K$ active suppliers, and their degrees are
		$$
			d(j) = \floor[\Big]{ \frac{v(K) - \mu w_j}{c} }
		$$
		where the value of $K$ is given in Theorem~\ref{thm:active-sups-at-equilibrium}.
	\label{thm:network-formation-nash-equilibrium-char}
\end{theorem}

\section{Price and Network Formation with Strategic Suppliers}
\label{sec:price-and-net-formation-game}

We now allow the suppliers to strategically choose prices,
and, then, retailers will form links in response. As before,
we will be interested in pure strategy Nash equilibria of this two-stage game.

\subsection{Price and Network Formation Game}

\begin{figure}
	\FIGURE
	{\includegraphics[width=0.2\linewidth]{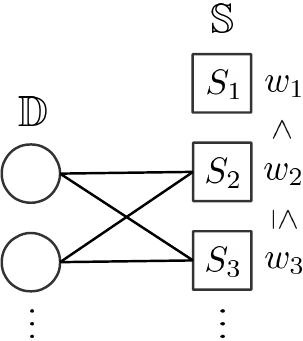}}
	{An example of a counterintuitive equilibrium network. \label{fig:weird-equil}}
	{
		Supply parameters are $\mu = 2$, $\sigma = 1$, $\Delta = 18$;
		linking cost is $c = 1/2$; supplier prices are $(w_1, w_2, w_3, \dots) = (12, 13, 13, \dots)$.
		In this equilibrium with $K = 2$ active suppliers,
		while the second and third
		suppliers are ``saturated'', with $d(2) = d(3) = \floor{(v(K) - \mu w_2) / c} = 2$,
		the marginal benefit of linking to the first supplier is negative: it is $-7/2$ if
		a vacant retailer links to it; and it is $-3/2$ if an
		active retailer changes one of its links $(i, 2)$ or $(i, 3)$ to $(i, 1)$.
	}
\end{figure}

For a given price vector $w$, there are many equilibrium networks. This is true at the very least
because for a given price vector $w$, links can be distributed almost arbitrarily among
the retailers, because an equilibrium is largely characterized by the number of active
suppliers and each supplier's degree. In fact, the network formation game with fixed wholesale
prices may possess counter-intuitive equilibria, in which expensive suppliers are
active, while the cheapest suppliers have no links, as Fig.~\ref{fig:weird-equil}
demonstrates.
We eliminate these equilibria with the following selection criterion.

\begin{definition}[Equilibrium Selection]
	For a vector $w \in \mathbb{R}_+^m$, let us define
	$\xi(w)$ to be the subset of pure strategy Nash equilibria---characterized
	in Theorem~\ref{thm:network-formation-nash-equilibrium-char}---in which the active
	suppliers have the lowest prices
	$$
		g^* \in \xi(w) \to \forall j \in \mathbb{S}_0(g^*): w_j \geq \max\limits_{\ell \in \mathbb{S}_+(g^*)}{w_\ell}.
	$$	
	Assume that,
	from the point of view of a supplier, all equilibria in $\xi(w)$ are
	equiprobable.
	\label{def:sets-of-network-equilibria}
\end{definition}

We, now, define the payoff of a supplier.

\begin{definition}[Payoff of a Supplier]
The payoff function
    of supplier $j$ is
    \begin{align}
        u_j(w) = a_j(w) S_j w_j,
        \label{eq:utility-supply-node}
    \end{align}
    where, if suppliers strategically choose prices $w$,
    $$
        a_j(w) = \pr[d(j, g^*) > 0 \mid g^* \in \xi(w)]
    $$
    is the likelihood that supplier $j$ is active in an equilibrium network
    $g^* \in \xi(w)$ subsequently formed by retailers in response to
    prices $w = (w_j, w_{-j})$ announced by suppliers; or, if the central
    planner decides upon prices $w$,
    $$
        a_j(w) = \delta\{d(j, g) > 0\},
    $$
    where $g$ is the network chosen by the central planner, and $\delta\{ \cdot \}$ is
    Kronecker's delta.
	\label{def:supply-node-utility}
\end{definition}

Supplier $j$'s payoff is computed under the assumption that the supplier sells its full
supply $S_j$
at price $w_j$ as long as it has at least one link in the network formed by retailers.
The latter may or may not happen, depending on which equilibria retailers
arrive at under $w$, or which network is chosen by
the central planner.

\begin{lemma}[Expected Payoff of a Supplier]
	\begin{align}	
		\E[u_j(w)] = a_j(w) \mu w_j,
		\label{eq:expected-supply-node-utility}
	\end{align}
	where $a_j(w)$ is the likelihood of supplier $j$'s being active in an equilibrium network,
	as per Definition~\ref{def:supply-node-utility}.
	\label{thm:expected-supply-node-utility}
\end{lemma}
\proof{Proof of Lemma~\ref{thm:expected-supply-node-utility}:}
	Trivially follows from equation~\eqref{eq:utility-supply-node} in
	Definition~\ref{def:supply-node-utility}. \qed
\endproof

\begin{definition}[Supply Chain Formation Game]
	A supply chain network formation game is a
	two-stage game, where, at the first stage, suppliers  announce prices $w$,
	maximizing their expected payoffs~\eqref{eq:expected-supply-node-utility}; and at the
	second stage, retailers play the network formation game with fixed prices $w$,
	as per Definition~\ref{def:network-formation-game}. A pure strategy Nash equilibrium price
	vector $w^*$ is a vector of prices immune to unilateral price deviations w.r.t. expected
	payoffs~\eqref{eq:expected-supply-node-utility}.  A pure strategy Nash equilibrium of the
	two-stage game is any pair $(w^*, g^*)$, where $g^* \in \xi(w^*)$, and $\xi(w^*)$ is
	characterized in Definition~\ref{def:sets-of-network-equilibria}.
	\label{def:price-and-network-formation-game}
\end{definition}

\subsection{Central Planner}

We define social welfare for the  two-stage supply chain formation game.

\begin{definition}[Social Welfare]
	The two-stage supply chain formation game, given in Definition~\ref{def:price-and-network-formation-game}, has the following social welfare
    \begin{align}
        Wel&fare(g, w) =
            \sum\limits_{i \in \mathbb{D}}{u_i(g, w)} +
            \sum\limits_{j \in \mathbb{S}}{u_j(g, w)} 
            + \int_{0}^{T(S)}{(\Delta - x)} \dif x
            - \sum\limits_{k \in \mathbb{S}_+(g)}{
                S_k (\Delta - T(S)),
            }
        \label{eq:welfare}
    \end{align}
    where
    $
        T(S) = T(S_1, \dots, S_m) = \sum_{k \in \mathbb{S}_+(g)}{S_k}
    $
    is the total supply of all active suppliers.
	\label{def:social-welfare}
\end{definition}

In~\eqref{eq:welfare}, the first two summands correspond
to the welfare of retailers and suppliers, respectively, and the last two summands describe consumer
surplus. The latter reflects the benefit the consumers enjoy due to their being able to purchase
a unit of product (supply) at the market price $(\Delta - T(S))$ rather than at the maximum price
$\Delta$, and corresponds to the area under the inverse demand curve above the market price.

\begin{lemma}[Expected Social Welfare]
	The expected social welfare in a two-stage supply chain formation game is as follows
    \begin{align}
        \E[Welfare] = 
            \mu K \Big(\Delta - \frac{\mu K}{2} \Big)
            - K \frac{\sigma^2}{2}
            - c |g|,
            \label{eq:expected-social-welfare}
    \end{align}
    where $K = K(g) = |\mathbb{S}_+(g)|$ is the number of active suppliers, and
    $|g|$ is the number of links in $g$.
	\label{thm:expected-social-welfare}
\end{lemma}
\proof{Proof of Lemma~\ref{thm:expected-social-welfare}:}
	As per~\eqref{eq:welfare}, expected social welfare consists of three components.
	
	The first component is the aggregate payoff of retailers:
    \begin{align*}
        \E\Big[\sum\limits_{i \in \mathbb{D}}{u_i(g, w)}\Big]
             &= \sum\limits_{i \in \mathbb{D}}{
                \sum\limits_{j \in N(i)}{
                    \left(
                        \frac{v(K) - \mu w_j}{d(j)} - c
                    \right)
                }
             }
		= \sum\limits_{j \in \mathbb{S}_+(g)}{
            \left(
                v(K) - \mu w_j - c d(j)
            \right)
        }\\
        &=K v(K) - \mu \sum\limits_{j \in \mathbb{S}_+(g)}{ w_j }  - c |g|,
    \end{align*}
    where the first equality comes from~\eqref{eq:expected-utility-retailer},
    and the last equality is valid because the double-summation is performed
    over all active suppliers, and each of them is counted in that double
    sum for every neighbor of $j$, that is, $d(j)$ times.
    
    The second component is the aggregate payoff of suppliers:
    \begin{align*}
        &\E\Big[\sum\limits_{j \in \mathbb{S}}{u_j(g, w)}\Big]
             = \mu \sum\limits_{j \in \mathbb{S}_+(g)}{
                a_j(w) w_j
             } = \mu \sum\limits_{j \in \mathbb{S}_+(g)}{
                w_j
             },
    \end{align*}
    which comes directly from equation~\eqref{eq:expected-supply-node-utility},
    and where the last equality is valid since every supplier $j$ we are summing
    over is active, so its likelihood $a_j(w)$ of being active is $1$.
   
   The third component is the expected consumer surplus:
    \begin{align*}
        &\E\Big[
            \int_{0}^{T(S)}{(\Delta - x)} \dif x
            - \sum\limits_{k \in \mathbb{S}_+(g)}{
                S_k (\Delta - T(S))
            }
        \Big]
        = (T(S) = \sum_{k \in \mathbb{S}_+(g)}{S_k})\\
		&=\E[(\Delta T(s) - \tfrac{1}{2} (T(S))^2) - (\Delta T(S) - (T(S))^2)]
		= \frac{1}{2} \E\Big[ T(S) T(S) \Big]\\[0.1in]
		&= \frac{1}{2} \E\Big[ (\sum_{j \in \mathbb{S}_+(g)}{S_j}) (\sum_{k \in \mathbb{S}_+(g)}{S_k}) \Big] = \frac{1}{2} (\sum_{j, k \in \mathbb{S}_+(g)}{\mu^2} + \sum_{j \in \mathbb{S}_+(g)}{\sigma^2})
			= \frac{K}{2}(\mu^2 K + \sigma^2).
    \end{align*}
    
    Gathering all three above components of expected social welfare, we get
    \begin{align*}
	    \E[Welfare] &= \sum_{j \in \mathbb{S}_+(g)}\Big(
		    		v(K) - \mu w_j - c d(j) + \mu w_j + \frac{K \mu^2 + \sigma^2}{2}
		    \Big)\\
		    &= \sum_{j \in \mathbb{S}_+(g)}\Big(
		    		\mu(\Delta - \mu K) - \sigma^2  - c d(j) + \frac{1}{2} (K \mu^2 + \sigma^2) \Big)\\
		    	&= \sum_{j \in \mathbb{S}_+(g)}\Big(
		    		\mu(\Delta - \frac{\mu K}{2}) - \frac{\sigma^2}{2}  - c d(j) \Big)
		    	= \mu K (\Delta - \frac{\mu K}{2}) - \frac{\sigma^2 K}{2}  - c |g|.
    \end{align*}
    \qed
\endproof

\begin{theorem}[Central Planner's Optimum]
	In a sufficiently large supply chain, an optimum of the central planner is a network
	where each of
    $$
        K^{opt} = \floor[\Big]{
			\frac{\Delta}{\mu} - \frac{1}{2} \left(\frac{\sigma}{\mu}\right)^2 - \frac{c}{\mu^2}
		} = \floor{y}
    $$
    active suppliers has exactly one link, and these links are distributed among retailers
    in an arbitrary fashion. The corresponding expected social welfare is
    \begin{align}
        \E[Welfare^{opt}]
			&= \frac{(\Delta \mu - \sigma^2 / 2 - c)^2- (\mu^2 \fracpart{y})^2}{2 \mu^2}.
        \label{eq:central-planner-opt-social-welfare}
    \end{align}
	\label{thm:central-planners-optimum}
\end{theorem}
\proof{Proof of Theorem~\ref{thm:central-planners-optimum}:}
	From Lemma~\ref{thm:expected-social-welfare}, we know that 
    \begin{align*}
        \E[Welfare] = 
            \mu K \Big(\Delta - \frac{\mu}{2}K \Big)
            - K \frac{\sigma^2}{2}
            - c |g| = h(K, |g|).
    \end{align*}
    First, we notice that, for the central planner, it does not make sense to
    have suppliers of degree larger than $1$, as raising the degree beyond $1$ would
    not affect $K$, and would only worsen the linking penalty term $- c |g|$. Thus,
    in an optimal solution, each supplier has exactly one link, resulting in the
    total number of links' matching the number of active suppliers, that is,
    $|g| = K$ (assuming that the supply chain is sufficiently large, with
    $n \geq K$). Thus, the central planner's solution is as follows:
    \begin{align*}
        h(K) &=
            \mu K \Big(\Delta - \frac{\mu}{2}K \Big)
            - K \frac{\sigma^2}{2}
            - c K \to \max,
     \end{align*}
     \begin{align*}
        h'(K) &= - \mu^2 K + \Delta \mu - \sigma^2 / 2 - c, \hence \\
        K^{opt} &=
		\floor[\Big]{
			\frac{\Delta}{\mu} - \frac{1}{2} \left(\frac{\sigma}{\mu}\right)^2 - \frac{c}{\mu^2}
        	 } = \floor{y} = y - \fracpart{y},\\
        h(K^{opt}) &= \frac{(\Delta \mu - \sigma^2 / 2 - c)^2 - (\mu^2 \fracpart{y})^2}{2 \mu^2}.
    \end{align*}
    Once again, we assume that the supply chain is sufficiently large, and $h$
    attains its maximum at $K^{opt}$ rather than at the boundary values $n$ or $m$.
    \qed
\endproof

\subsection{Nash Equilibrium Analysis}

We start by analyzing the behavior of suppliers---namely, the prices they set---at an
equilibrium of the two-stage game.

\begin{theorem}[Equilibrium Prices]
	In a sufficiently large two-stage supply chain formation game,
	at a pure strategy Nash equilibrium, $w^*= \zerobb$.
	\label{thm:price-nash-equilibrium-char}
\end{theorem}
\proof{Proof of Theorem~\ref{thm:price-nash-equilibrium-char}:}
	As before, we assume that the prices are ordered as $w_1 \leq w_2 \leq \dots w_m$.
	
	Suppliers' behavior is driven by their expected payoff~\eqref{eq:expected-supply-node-utility}
	$$
		\E[u_j(w)] = a_j(w) w_j \mu,
	$$
	where $\mu$ is constant, $w_j$ is the price chosen by supplier $j$, and $a_j(w)$ is the
	likelihood of supplier $j$'s being active in network equilibria potentially formed by retailers
	in response to the announced prices $w = (w_j, w_{-j})$.
	
	First, let us prove that, at an equilibrium, all the suppliers set identical prices, that is,
	$w^* = \const \cdot \onebb$.
	
	From Theorem~\ref{thm:active-sups-at-equilibrium}, we know that the number of active
	suppliers in an equilibrium network is either $K^*$ or $(K^* - 1)$, where
	$$
		K^* = \min\{ K \in \mathbb{Z}_+ \mid v(K + 1) - \mu w_{K + 1} - c < 0 \},
	$$
	and it can be only $K^*$ if $v(K^*) - \mu w_{K^*} - c < 0$ (in contrast to being exactly
	zero).
	
	If the latter expression is indeed negative, then prices $w_{K^* + 1}, \dots, w_m$ cannot
	be strictly larger than $w_{K^*}$ at an equilibrium (since, if they were, the corresponding suppliers
	would never be linked to, making their expected payoffs $0$), and prices $w_1, \dots, w_{K^* - 1}$
	cannot be bounded away from $w_{K^*}$ (as the corresponding suppliers will be linked at an
	equilibrium anyway, and they can increase their prices up to $(w_{K^*} - \epsilon)$ without
	affecting their likelihoods $a_j(w) = 1$ of being active).
	
	If $v(K^*) - \mu w_{K^*} - c = 0$, that is, if supplier $K^*$'s activation does not affect
	a retailer's expected payoff, reasoning similarly to the previous case,
	the prices $w_1, \dots, w_{K^* - 1}$ should be identical---it does not make sense to
	keep a price below $w_{K^* - 1}$, as the cheapest $(K^* - 1)$ suppliers will be linked
	to (assuming an existing gap between $w_{K^* - 1}$ and $w_{K^*}$). At the same time,
	$w_{K^*}, w_{K^* + 1}, \dots, w_m$ should also be identical, as setting a price
	larger than $w_{K^*}$ would make $a_j(w) = 0$.
	\begin{figure}[h!]
		\centering
		\includegraphics[width=0.75\linewidth]{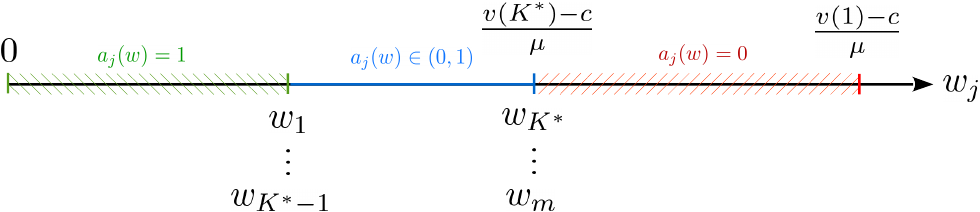}
	\end{figure}
	
	Now, the question is whether it is better to have $a_j(w) = 1$ and a smaller price $w_1$,
	or $a_j(w) \in (0, 1)$ and a larger price $w_{K^*}$. The likelihood $a_j(w)$ of being linked
	to for suppliers $K^*, \dots, m$ reflects how often one of these suppliers is chosen to be
	the $K^*$'th active supplier, and is inversely proportional to $m$. Recalling our assumption
	that $m$ is sufficiently large, we can conclude that $a_j(w) \in (0, 1)$ can be made arbitrarily
	small, making $a_j((w_1, w_{-1})) w_1 = w_1 > a_j((w_{K^*}, w_{-K^*})) w_{K^*}$, thereby,
	driving prices $w_{K^*}, \dots, w_m$ towards $w_1$.
	
	Thus, the suppliers have an incentive to set identical prices at an equilibrium.
	
	Finally, we notice that, if $w = \const \cdot \onebb$, then every supplier has an incentive
	to reduce
	its price by an infinitesimal amount, becoming the cheapest supplier and increasing its likelihood
	of being active from an arbitrarily small $a_j(w) \in (0, 1)$ (as the number of supplier competing on the price scales together with $m$ that can be arbitrarily large) to $a_j(w) = 1$. As a result,
	all the prices are driven towards their lower bound, which in this case is $0$.
	\qed
\endproof

According to Theorem~\ref{thm:price-nash-equilibrium-char}, at an equilibrium, suppliers trade
at a zero profit. This insight allows us to revisit our previous statements about the number of
active suppliers as well as their degrees at a network equilibrium, which we do in the following
theorem.

\begin{theorem}[Nash Equilibria in Two-stage Game]
	In a sufficiently large two-stage supply chain formation game (see Definition~\ref{def:price-and-network-formation-game}), at
	any pure strategy Nash equilibrium $(g^*, w^*)$ of this game, $w^* = \zerobb$, and 
	the number of active suppliers is either $K = K^*$, with
	\begin{align*}
		K^* = \floor[\Big]{
			\frac{\Delta}{\mu} - \left(\frac{\sigma}{\mu}\right)^2 - \frac{c}{\mu^2}
		} = \floor{z},
	\end{align*}
	or $K = K^*$ or $K = (K^* - 1)$ if $v(K^*) = c$. Each of the active suppliers has degree
	$
		d(j) = \floor{v(K) / c},
	$
	and, more specifically,
			if $K = K^*$, then 
			$
				d(j) = \floor{ 1 + \mu^2 \fracpart{z} / c }
			$;
			if $K = K^* - 1$, then
				$
					d(j) = \floor{1 + \mu^2 (1 + \fracpart{z}) / c};
				$
	where $K = K^* - 1$ is possible only if $v(K^*) = c$.
	\label{thm:two-stage-game-equil}
\end{theorem}
\proof{Proof of Theorem~\ref{thm:two-stage-game-equil}:}
	Applying Theorem~\ref{thm:active-sups-at-equilibrium} to the case $w^* = \zerobb$, we
	immediately get
	\begin{align*}
		K^* &= \min\{ K \in \mathbb{Z}_+ \mid v(K + 1) - c < 0 \},
	\end{align*}
	where $v(K) = \mu(\Delta - \mu K) - \sigma^2$, so
	\begin{align*}
			K^* &= \floor[\Big]{
			\frac{\Delta}{\mu} - \left(\frac{\sigma}{\mu}\right)^2 - \frac{c}{\mu^2}
		} = \floor{ z }.
	\end{align*}
	Furthermore, from Theorem~\ref{thm:network-formation-nash-equilibrium-char}, we have
	$$
		d(j) = \floor{v(K) / c}.
	$$
	
	\noindent If $K = K^*$:
	$
		d(j) = \floor{v(K^*) / c} = \floor{v(\floor{z}) / c}
			= \floor{1 + \mu^2 \fracpart{z} / c}.
	$
	
	\vspace{0.1in}
	\noindent If $K = K^* - 1$:
	$
		d(j) = \floor{v(K^* - 1) / c} = \floor{v(\floor{z} - 1) / c}
			= \floor{1 + \mu^2 (1 + \fracpart{z}) / c}.
	$
	\qed
\endproof

In the analysis of equilibrium efficiency, we consider only the case when $v(K^*) > c$ and,
hence, the number of active suppliers in every equilibrium is exactly $K^*$. The analysis for
the case of $v(K^*) = c$ and $K = K^*$ or $K = K^* - 1$ is very similar, and brings no
additional insights.

\begin{theorem}[Equilibrium Welfare]
	In a sufficiently large two-stage supply chain formation game
	(see Definition~\ref{def:price-and-network-formation-game}), 
 its pure strategy Nash equilibria, characterized in Theorem~\ref{thm:two-stage-game-equil},
	have the following expected social welfare
	\begin{align*}
		\E[Welfare]
			=
			\frac{
				(\Delta \mu - c - \mu^2\fracpart{z} - \sigma^2)
				(\Delta \mu - c - \mu^2 \fracpart{z} + 2c\fracpart{\mu^2 \fracpart{z} / c})
			}{
				2 \mu^2
			},
	\end{align*}
	where $z = \frac{\Delta}{\mu} - \left(\frac{\sigma}{\mu}\right)^2 - \frac{c}{\mu^2}$.
	\label{thm:nash-equilibrium-efficiency}
\end{theorem}
\proof{Proof of Theorem~\ref{thm:nash-equilibrium-efficiency}:}
	From Theorem~\ref{thm:two-stage-game-equil}, we have expressions for both the
	number $K^*$ of active suppliers, and their degrees $d(j)$ at an equilibrium $g^*$:
	\begin{align*}
		K^* &= \floor[\Big]{
			\frac{\Delta}{\mu} - \left( \frac{\sigma}{\mu} \right)^2 - \frac{c}{\mu^2}
		} = \floor{z} = z - \fracpart{z},\\
		d(j) &= \floor{ 1 + \mu^2 \fracpart{z} / c },
	\end{align*}
	where $j \in \mathbb{S}_+(g^*)$. As every supplier has the same degree, then the total
	number of links is
	$$
		|g| = K d(j) = K \floor{ 1 + \mu^2 \fracpart{z} / c }.
	$$
	Substituting three above expressions in the expression~\eqref{eq:expected-social-welfare}
	\begin{align*}
		\E[&Welfare] =
            \mu K \Big(\Delta - \frac{\mu K}{2} \Big)
            - K \frac{\sigma^2}{2}
            - c |g|
	\end{align*}
	for expected social welfare, we get the expression in the theorem's statement.
	\qed
\endproof

\section{Price Formation Under Heterogeneous Supply Variance}
\label{sec:prices-with-varying-sigma}

In this section, we
allow supply variance  to vary across suppliers. For simplicity, we will consider a small deviation from the case of identical
suppliers by allowing the first supplier to be strictly more reliable. For this case, the price formation behavior of
the supplier is characterized in the following theorem.

\begin{theorem}[Prices at Equilibrium with Heterogeneous Supply Variance]
	In a two-stage supply chain formation game with a sufficiently large number of strategic
	retailers and suppliers, if random supplies have identical means $\E[S_j] = \mu$ and
	non-identical variances $\Var[S_1] = \sigma_1^2 < \sigma^2_2 = \Var[S_2] = \dots = \Var[S_m]$,
	and suppliers perform equilibrium selection ignoring equilibria where ``high-value'' suppliers
	are not linked to
	$$
		g^* \in \xi(w) \to \forall j \in \mathbb{S}_0(g^*): \sigma^2_j + \mu w_j \geq \max\limits_{\ell \in \mathbb{S}_+(g^*)}{\{\sigma^2_\ell + \mu w_\ell\}}.
	$$	
	then, at a pure strategy Nash equilibrium,
	$$
		w^*_1 = \frac{\sigma_2^2 - \sigma_1^2}{\mu} - \varepsilon,\ w^*_2 = \dots  = w^*_m = 0,
	$$
	where $\varepsilon$ is a positive real value approaching $0$.
	\label{thm:price-nash-equilibrium-char-varying-sigma}
\end{theorem}
\proof{Proof of Theorem~\ref{thm:price-nash-equilibrium-char-varying-sigma}:}
Following Lemma~\ref{thm:expected-retailer-utility}, we compute the expected payoff
of a retailer under fixed $w$ as
\begin{align*}
	\E[u_i(g, w)] = \sum\limits_{j \in N(i)}{\Big(
		\frac{v_j(K) - \mu w_j}{d(j)} - c
	\Big)},
	\quad
	v_j(K) = \mu (\Delta - \mu K) - \sigma_j^2,
	K = K(g) = |\mathbb{S}_+(g)|.
\end{align*}

Algorithm~\ref{alg:greedy-equil-gen} that greedily constructs a pure strategy Nash equilibrium
network with the largest number $K_{max}^*$ of active suppliers at an equilibrium still applies
to this case, except that the algorithm
now selects suppliers having ordered them in an ascending order by $(\sigma^2_j + \mu w_j)$
rather than by $w_j$ in the case of identical suppliers. With that change, with get an analog of Theorem~\ref{thm:network-formation-equilibrium-existence} stating equilibrium existence, and an analog of Theorem~\ref{thm:active-sups-at-equilibrium}
that establishes the number of active suppliers at an equilibrium.

Analogously to Theorem~\ref{thm:network-formation-nash-equilibrium-char}, we establish that
supplier $j$ has the following degree at an equilibrium is
$
	d(j) = \floor{
		(v_j(K) - \mu w_j) / c
	}
$.

Finally, price formation happens similarly to how it is described in the proof of
Theorem~\ref{thm:price-nash-equilibrium-char}, with one qualitative difference. While the prices
are driven towards $0$, the first supplier---which has a strictly lower supply variance $\sigma_1^2$---has
advantage over other suppliers having a higher supply variance $\sigma^2_2$. For supplier 1 and
any other supplier, say, 2, to be equivalent from the point of view of retailers linking to them,
it must be that $\sigma_1^2 + \mu w_1 = \sigma_2^2 + \mu w_2$ or, equivalently,
$
	w_1 - w_2 = (\sigma_2^2 - \sigma^2_1) / \mu
$.
Consequently, while $w_2$ is driven to $0$, the first supplier can set its price $w_1^*$ to any value
below $(\sigma_2^2 - \sigma^2_1) / \mu$, thereby, guaranteeing itself the status of the
``best-value'' supplier that is always linked to in the considered equilibria $\xi(w)$, making
its expected payoff
$$
	\E[u_j(w)] = a_j(w) \mu w_1^* = \mu w_1^* \approx \sigma_2^2 - \sigma_1^2.
$$
If, instead, the first supplier decided to set its price to a value strictly larger than
$(\sigma_2^2 - \sigma^2_1) / \mu$, then it would never be linked to by the retailers, making
its expected payoff $0$. If it set its price to exactly $(\sigma_2^2 - \sigma^2_1) / \mu$,
then, from the retailers' perspective, it would be equivalent to all the other suppliers, whose
number $(m - 1)$ can be arbitrarily large and, consequently, the first supplier's likelihood
$a_j(w)$ of being linked to would be arbitrarily small, as would be that supplier's expected
payoff.
\qed
\endproof

Theorem~\ref{thm:price-nash-equilibrium-char-varying-sigma} establishes that suppliers
have an incentive to improve their reliability, as the latter would allow them to trade at a positive marginal
profit, in contrast to the zero marginal profit when all suppliers are identical; and the value of the
marginal profit is determined by the difference in supply variances.

Having stated the result for prices at equilibrium, we will now characterize how the improvement
in a supplier's supply variance affects social welfare, ignoring the infinitesimal $\varepsilon$ in
the expression for the price of the first (improved) supplier obtained in Theorem~\ref{thm:price-nash-equilibrium-char-varying-sigma}.
\begin{theorem}
	Under the conditions of Theorem~\ref{thm:price-nash-equilibrium-char-varying-sigma}, when
	one supplier has a strictly better supply variance $\sigma_1^2 < \sigma^2_2 = \dots = \sigma^2_m$, the total expected social welfare changes as
	$$
		\E[Welfare] = \E[Welfare_{ident}] + \tfrac{1}{2} (\sigma_2^2 - \sigma_1^2),
	$$
	where $\E[Welfare_{ident}]$ is the social welfare for the supply chain with identical
	suppliers, characterized in Theorem~\ref{thm:nash-equilibrium-efficiency}.
	In particular,
	\begin{enumerate}
		\item the welfare of suppliers increases by $\E[u_1(w^*)] = \sigma_2^2 - \sigma_1^2$;
		\item the welfare of retailers is unchanged; and
		\item consumers' welfare decreases by $\tfrac{1}{2}(\sigma_2^2 - \sigma_1^2)$.
	\end{enumerate}
	\label{thm:social-welfare-varying-sigma}
\end{theorem}
\proof{Proof of Theorem~\ref{thm:social-welfare-varying-sigma}:}
	First, notice that, when the first supplier improves its supply variance, this does not affect
	either the number $K^*$ of active suppliers or the degree $d(j)$ of every supplier at
	an equilibrium. Indeed, $K^*$ is defined as
	$$
		K^* = \min\{ K \in \mathbb{Z}_+ \mid v_{K + 1}(K + 1) - \mu w_{K + 1} - c < 0 \}.
	$$
	$K$ cannot be $0$, as the opposite would violate Assumption~\ref{asm:bounded-prices-and-linking-costs} about each supplier's ``value'' being non-negative in the absence of links. This,
	however, mean that the expression
	$$
		K^* = \floor[\Big]{
			\frac{\Delta}{\mu} - \left(\frac{\sigma}{\mu}\right)^2 - \frac{c}{\mu^2}
		} = \floor{z}
	$$
	for $K^*$ from Theorem~\ref{thm:two-stage-game-equil} is still valid even when the
	first supplier changes its supply variance. Furthermore, at an equilibrium
	$$
		v_1(K) - \mu w_1^* = \mu(\Delta - \mu K) - \sigma_1^2 - \mu w_1^*
			= \mu(\Delta - \mu K) - \sigma_1^2 - \mu \frac{\sigma_2^2 - \sigma_1^2}{\mu}
			= \mu(\Delta - \mu K) - \sigma_j^2 = v_j(K) - \mu w_j^*,
	$$	
	where $j > 1$, so the previously derived expression for the supplier degree at an equilibrium
	$
	d(j) = \floor{
		(v_j(K) - \mu w_j) / c
	}
	$
	is still valid, and according to it, all suppliers still have the same degree at an equilibrium.
	
	Equipped with the two observations above, we can now follow the computation of expected
	social welfare from Theorem~\ref{thm:expected-social-welfare}, and analyze what happens
	to its different components when the first supplier improves its supply variance.
	
	The retailers' welfare, defined as
	\begin{align*}
		\E\Big[\sum\limits_{i \in \mathbb{D}}{u_i(g, w)}\Big]
             &= \sum\limits_{i \in \mathbb{D}}{
                \sum\limits_{j \in N(i)}{
                    \left(
                        \frac{v_j(K) - \mu w_j}{d(j)} - c
                    \right)
                }
             }
	\end{align*}
	clearly does not change as a result of the change in $\sigma_1^2$, as both
	$v_j(K) - \mu w_j = \const$ and $d(j) = \const$ across all the suppliers at an equilibrium.
	
	Suppliers' welfare
	\begin{align*}
	 &\E\Big[\sum\limits_{j \in \mathbb{S}}{u_j(g, w)}\Big]
             = \mu \sum\limits_{j \in \mathbb{S}_+(g)}{
                w_j^*
             } = \mu w_1^* = \mu \frac{\sigma_2^2 - \sigma_1^2}{\mu} = \sigma_2^2 - \sigma_1^2
	\end{align*}
	includes zero welfare of the majority of the suppliers who set zero prices, and
	positive welfare $\sigma_2^2 - \sigma_1^2$ of the first supplier, while it is used to
	be $0$ in the case of identical suppliers.
	
	Finally, consumer surplus changes as
	\begin{align*}
		\E\Big[&
            \int_{0}^{T(S)}{(\Delta - x)} \dif x
            - \sum\limits_{k \in \mathbb{S}_+(g)}{
                S_k (\Delta - T(S))
            }
        \Big]
		=
		\frac{1}{2} \E\Big[ (\sum_{j \in \mathbb{S}_+(g)}{S_j}) (\sum_{k \in \mathbb{S}_+(g)}{S_k}) \Big]
		\\[0.1in]
		&=
		\frac{1}{2} ( \sum_{j, k \in \mathbb{S}_+(g)}{\mu^2} + \sum_{j \in \mathbb{S}_+(g)}{\sigma_j^2} )
		= \frac{1}{2} (K^2 \mu^2 + (K - 1) \sigma_2^2 + \sigma_1^2)
		= \frac{1}{2} (K^2 \mu^2 + K \sigma_2^2 + \sigma_1^2 - \sigma_2^2)\\[0.1in]
		 &= \frac{K}{2} (K \mu^2 + \sigma_2^2) - \frac{\sigma_2^2 - \sigma_1^2}{2}
		 = \E[ConsumerWelfare_{ident}] - \frac{\sigma_2^2 - \sigma_1^2}{2},
	\end{align*}
	where $\E[ConsumerWelfare_{ident}]$ is expected consumer surplus for the case of
	identical suppliers, calculated in the proof of Lemma~\ref{thm:expected-social-welfare}.
	
	If we collect the changes to welfare of suppliers, retailers, and consumers above, we will
	arrive at the conclusion that the total expected social welfare---being the sum of the three
	above mentioned components---increases by $(\sigma_2^2 - \sigma_1^2) / 2$, with
	consumers paying that amount, and suppliers earning twice that much.
	\qed
\endproof

According to Theorem~\ref{thm:social-welfare-varying-sigma}, improvement of supplier reliability
benefits the corresponding suppliers, while retailers are not being affected, and the consumers
face the reliability improvement cost.

\section{Price Formation Under Heterogeneous Supply Expectation}
\label{sec:prices-with-varying-mu}

While in the previous section, we established that suppliers are incentivized to improve their
reliability to make positive marginal profit, the natural question is whether an analogous statement
about improving expected supply is also valid.

Let us consider a simple environment similar to the one in the previous section, but let the first
supplier to have a strictly better expected supply
$$
	\mu_1 = \mu + \delta, (\delta > 0), \quad \mu_2 = \mu_3 = \dots = \mu_m = \mu,
$$
while all the suppliers are identical w.r.t. supply variance $\sigma^2$.

\begin{theorem}[Network Equilibria with Heterogeneous Supply Mean]
	In a sufficiently large supply chain network formation game with fixed $w$, where suppliers have
	identical supply variances $\Var[S_j] = \sigma^2$, yet, the first supplier has a strictly better mean supply
	$
		\E[S_1] = \mu + \delta > \mu = \E[S_2] = \E[S_3] = \dots = \E[S_m],
	$
	let us put
	\begin{align*}
		B(K) &= \Big\{
				s \subseteq \mathbb{S} \mid
				|s| = K;
				\forall j \in s: v_j(s) - \mu_j w_j - c \geq 0;\\
				&\forall s' \neq s: |s'| = |s| \to \sum\limits_{j' \in s'}{(\Delta - \sum_{\ell \in s'}{\mu_\ell} - w_{j'})\mu_{j'}}
					\leq \sum\limits_{j \in s}{(\Delta - \sum_{\ell \in s}{\mu_\ell} - w_{j})\mu_{j}}
		\Big\}
	\end{align*}
	to be the set of cardinality-$K$ subsets of best suppliers, with $v_j(s) = (\Delta - \sum_{\ell \in s}{\mu_\ell})\mu_j - \sigma^2$. Then
	\begin{itemize}
		\vspace{0.05in}
		\item{
			a pure strategy Nash equilibrium of that game exists; and
		}
		\vspace{0.05in}
		\item{
			the largest number of active suppliers in an equilibrium network is
			\begin{align*}
				K_{max}^* = \min\{
					K \mid \forall K^+ > K : B(K^+) = \emptyset
				\}.
			\end{align*}
		}
	\end{itemize}	
	\label{thm:equil-networks-when-varying-mu}
\end{theorem}
\proof{Proof of Theorem~\ref{thm:equil-networks-when-varying-mu}:}
	The definition of the largest number $K_{max}^*$ of active suppliers at an equilibrium
	together with the greedy construction of an equilibrium with such number of active suppliers
	goes along the lines of Algorithm~\ref{alg:greedy-equil-gen} and
	Lemma~\ref{thm:greedy-equil-constr}---we, first, activate as many suppliers as possible and,
	subsequently, attach as many links as possible to every active supplier using a vacant demand
	node for every link creation. However, there is one difference. Here, we cannot rank suppliers
	by price $w_j$ any longer, and, furthermore, there is no static ranking of suppliers. As a result,
	we define sets $s \in B(K)$ of best suppliers of size $K$ and pick one of them---that
	corresponds to the largest $K_{max}^*$---for an equilibrium obtained by attaching as many
	links as possible to the chosen suppliers. Notice, that $K_{max}^*$ is
	well-defined, as $v_j(s)$ monotonically decreases in the number of active suppliers $|s|$,
	while the total number $m$ of suppliers is sufficiently large, and, at 	some point, condition
	$v_j(s) - \mu_j w_j - c \geq 0$ will not hold for any supplier in the system.
	
	Existence of equilibrium immediately follows, as in
	Theorem~\ref{thm:network-formation-equilibrium-existence} for the case of identical suppliers.
	\qed
\endproof

\begin{theorem}[Prices at Equilibrium with Heterogeneous Supply Mean]
	In a two-stage supply chain formation game with a sufficiently large number of strategic
	retailers and suppliers, if random supplies have identical variances $\Var[S_j] = \sigma^2$ and
	non-identical means $\mu_1 = \mu + \delta > \mu = \E[S_2] = \dots = \E[S_m]$,
	and suppliers perform equilibrium selection ignoring equilibria where ``high-value'' suppliers
	are not linked to
	$$
		g^* \in \xi(w) \to \forall j_0 \in \mathbb{S}_0(g^*):
		\Big(\Delta - \sum\limits_{\ell \in \mathbb{S}_+(g^*)}{\mu_\ell} - w_{j_0}\Big) \mu_{j_0}
		\geq 
		\max\limits_{j_+ \in \mathbb{S}_+(g^*)}{\Big\{
			\Big(\Delta - \sum\limits_{\ell \in \mathbb{S}_+(g^*)}{\mu_\ell} - w_{j_+}\Big) \mu_{j_+}
		\Big\}}.
	$$	
	if the largest number $K_{max}^*$ of active suppliers at an equilibrium, defined in
	Theorem~\ref{thm:equil-networks-when-varying-mu}, is greater than 1, then
	at a pure strategy Nash equilibrium, supplier prices are
	$$
		w^*_1 = \delta \left(
			\frac{\Delta - \mu K_{max}^*}{\mu + \delta} - 1
		\right) - \varepsilon,\
		w^*_2 = \dots  = w^*_m = 0,
	$$
	where $\varepsilon > 0$ approaches $0$.
	If $K_{max}^* = 1$, then $(w_2^*, \dots, w_m^*) \in [0; \Delta - \mu - (\sigma^2 + c) / \mu]^{m - 1}$.
	\label{thm:price-nash-equilibrium-char-varying-mu}
\end{theorem}
\proof{Proof of Theorem~\ref{thm:price-nash-equilibrium-char-varying-mu}:}
	Reasoning from the proof of Theorem~\ref{thm:price-nash-equilibrium-char} entails that identical
	suppliers $2, 3, \dots, m$ are driven towards setting identical prices, and their willingness to boost
	their likelihoods $a_j(w)$ of being linked to at a network equilibrium from a sufficiently small
	value in $(0, 1)$ to the value of $1$ drives the prices to $0$.
	\begin{figure}[h!]
		\centering
		\includegraphics[width=0.6\linewidth]{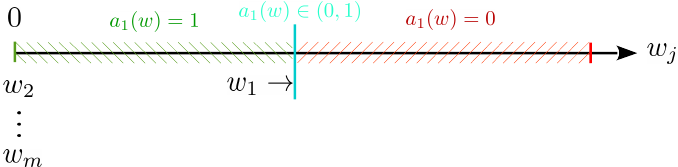}
	\end{figure}
	
	\noindent In this setting, if we assume a particular number $K$ of active suppliers at
	an equilibrium, if the first supplier set its price $w_1$ to be
	$$
		w_1 = \delta \left(
			\frac{\Delta - \mu K}{\mu + \delta} - 1
		\right),
	$$
	it would entail 
	$
	\forall j > 1: (\Delta - \sum_{\ell \in \mathbb{S}_+(g^*)}{\mu_\ell} - w_{j}) \mu_{j}
		 = (\Delta - \sum_{\ell \in \mathbb{S}_+(g^*)}{\mu_\ell} - w_{1}) \mu_{1},
	$
	that is, the first supplier would have been indistinguishable from the rest of the suppliers
	from the perspective of a retailer. If supplier uses such price, then it will compete with a
	number of suppliers that scales together with the number $(m - 1)$ of other suppliers, making
	its likelihood 	$a_1(w)$ of being linked to an arbitrarily small value
	(as the size $m$ of the chain is sufficiently large). Furthermore, as all the other suppliers
	set identical prices, supplier $1$ will 
	be indistinguishable from either all or none of them. Hence, to boost its likelihood $a_1(w)$ of
	being linked to from an arbitrarily small value to $1$, the first supplier needs to make
	sure that regardless of what the number $K$ of active suppliers at an equilibrium is,
	this supplier's ``value'' is
	strictly higher than that of every other supplier. Consequently, the price of this supplier
	approaches the price at which it is indistinguishable from other suppliers at an equilibrium
	with the largest number of active suppliers from the left
	$$
		w_1^* \to \delta \left(
			\frac{\Delta - \mu K_{max}^*}{\mu + \delta} - 1
		\right) - 0.
	$$
	
	If, however, it happens that $K_{max}^* = 1$, then, while the first supplier conservatively
	sets its price as described above and gets links, all the other suppliers will have no links
	regardless of their prices, and, hence, $(w_2^*, \dots, w_m^*) \in [0; \Delta - \mu - \frac{\sigma^2 + c}{\mu}]^{m - 1}$, where the upper bound on the supplier price comes from Assumption~\ref{asm:bounded-prices-and-linking-costs}.
	\qed
\endproof

Having established how suppliers set prices at an equilibrium, we can characterize how social
welfare changes in response to the first supplier's improving its capacity. In what follows, we will
ignore the corner-case $K_{max}^* = 1$ from Theorem~\ref{thm:price-nash-equilibrium-char-varying-mu}, that corresponds to a large number of non-informative equilibria.

\begin{theorem}
	Under the conditions of Theorem~\ref{thm:price-nash-equilibrium-char-varying-mu}, when
	one supplier has a strictly higher mean supply $\mu_1 = \mu + \delta > \mu = \mu_2 = \dots = \mu_m$,
	\begin{itemize}
		\item the \emph{total} expected welfare \emph{increases} by approximately
			$\delta (\Delta - \mu (K_{max}^* - K^* + 1) - \delta / 2) > 0$
		\item the expected welfare of \emph{suppliers increases} by
			$\delta (\Delta - \mu (K^*_{max} + 1) - \delta) > 0$;
		\item the expected welfare of \emph{consumers increases} by
			$\delta (\mu K^* + \delta / 2) > 0$;
		\item while \emph{retailers'} expected welfare approximately \emph{does not change};
	\end{itemize}
	where $K^*$ is the number of active suppliers in an equilibrium network formed by the retailers,
	and $K_{max}^*$ is the largest number of active suppliers at an equilibrium.
	\label{thm:social-welfare-varying-mu}
\end{theorem}
\proof{Proof of Theorem~\ref{thm:social-welfare-varying-mu}:}
	From Theorem~\ref{thm:price-nash-equilibrium-char-varying-mu}, we know 
	\begin{align*}
		w_1^* \approx \delta \left(
			\frac{\Delta - \mu K_{max}^*}{\mu + \delta} - 1
		\right),
		w_j^* = 0, j > 1.
	\end{align*}
	Now, we can substitute these prices, together with $\mu_1 = \mu + \delta, \mu_j = \mu, j > 1$
	into the expressions for expected welfare of suppliers, retailers, and consumers, coming from
	equation~\eqref{eq:welfare} in Definition~\ref{def:social-welfare}. (In what follows, notation
	$X'$ will mean $X$ after the first supplier increased its mean supply.)
	
	Suppliers' welfare changes as
	\begin{align*}
		\E[SupplierWelfare'] &= \E[SupplierWelfare] + \mu_1 w^*_1\\
			&= \E[SupplierWelfare] + \delta (\Delta - \mu (K_{max}^* + 1) - \delta),
	\end{align*}
	where $\delta \Delta$ is the dominant term scaling together with the total number $m$ of
	suppliers, making the second summand in the obtained expression positive. Thus,
	the welfare of suppliers (actually, just the welfare of the first supplier) increases.
	
	Consumers' surplus changes as
	\begin{align*}
		\E[&ConsumerWelfare'] = \E\Big[
            \int_{0}^{T(S)}{(\Delta - x)} \dif x
            - \sum\limits_{k \in \mathbb{S}_+(g^*)}{
                S_k (\Delta - T(S))
            }
        \Big]\\
        &= \frac{1}{2} \E\Big[ (\sum_{j \in \mathbb{S}_+(g^*)}{S_j}) (\sum_{k \in \mathbb{S}_+(g^*)}{S_k}) \Big]
		= \frac{1}{2} \E\Big[ S_1^2 + 2 S_1 \sum_{j \in \mathbb{S}+ \setminusx \{1\}}{S_j} + (\sum_{j \in \mathbb{S}+ \setminusx \{1\}}{S_j})^2 \Big]\\
		&= \frac{1}{2} \Big[ (\mu + \delta)^2 + \sigma^2 + 2 (\mu + \delta) (K^* - 1) \mu + (K^* - 1)^2 \mu^2 + (K^* - 1) \sigma^2 \Big]  \\[0.1in]
		&= \frac{1}{2} \Big[   
			(\mu^2 + 2(K^* - 1) \mu^2 + (K^* - 1)^2\mu^2)
			+ K^* \sigma^2
			+ (2\mu\delta + \delta^2 + 2 (K^* - 1) \mu \delta)
		\Big]\\[0.1in]
        &= \frac{1}{2} [ {K^*}^2 \mu^2 + {K^*} \sigma^2 ] + \frac{1}{2} (2 K^* \mu \delta + \delta^2)
        		=  \E[ConsumerWelfare] + \delta (K^* \mu + \delta / 2).
	\end{align*}
	In the obtained expression, $K^*$ is the number of active suppliers in a particular equilibrium
	network formed by the retailers that, generally, can be smaller than $K_{max}^*$.
	
	Prior to computing the change in retailer and total welfare, we need to establish how supplier
	degrees at equilibrium change after the first supplier increases its mean supply. As before the
	degree of supplier $j$ at equilibrium is
	\begin{align*}
		d(j, g^*) = \floor[\Big]{
			\frac{ v_j(g^*) - \mu_j w_j^* }{c}
		} = \floor[\Big]{
			\Big((\Delta - \sum\limits_{\ell \in \mathbb{S}_+(g^*)}{\mu_\ell})\mu_j - \sigma^2 - \mu_j w_j^* \Big) / c
		}.
	\end{align*}
	Ignoring the small $\varepsilon$ in the expression for $w_1^*$, and substituting that price into
	the above expression for a supplier's degree, we obtain
	\begin{align*}
		d(1)' &= \floor[\Big]{
			\frac{(\Delta - K^* \mu) \mu - \sigma^2}{c} +
			\frac{\delta \mu (K_{max}^* - K^* + 1)}{c}
		} \approx \floor[\Big]{
			\frac{(\Delta - K^* \mu) \mu - \sigma^2}{c}
		} + \floor[\Big]{
			\frac{\delta \mu (K_{max}^* - K^* + 1)}{c}
		}\\
		&= d(1) + \floor[\Big]{
			\frac{\delta \mu (K_{max}^* - K^* + 1)}{c}
		}.
	\end{align*}
	Doing the same for $d(j)$, $j > 1$, we obtain
	\begin{align*}
		d(j)' = \floor[\Big]{
			\frac{(\Delta - K^* \mu)\mu - \sigma^2 - \mu w_j^*}{c} - \frac{\delta \mu}{c}
		} \approx d(j) - \ceil[\Big]{ \frac{\delta \mu}{c} }.
	\end{align*}
	Now, we can use the derived above expressions for supplier degrees to compute the change
	to retailer welfare.
	\begin{align*}
		\E[&RetailerWelfare]' = \sum\limits_{j \in \mathbb{S}_+(g^*)}{\left( v_j(g)' - \mu_j w_j^* - c d(j)' \right)}\\
			 &= (v_1(g)' - (\mu + \delta) w_1^* - c d(1)') + \sum\limits_{j \in \mathbb{S}_+(g^*) \setminusx \{1\}}{\left( v_j(g) - \delta \mu - \mu w_j^* - c d(j)' \right)}\\
			 &= (\Delta - K^* \mu - \delta) (\mu + \delta) - \sigma^2 - c d(1) - \delta (\Delta - \mu K_{max}^* - \mu - \delta) - c \floor[\Big]{\frac{\delta \mu}{c}(K_{max}^* - K^* + 1)}\\
			&\phantom{xxx}
			+ \sum\limits_{j \in \mathbb{S}_+(g^*) \setminusx \{1\}}{\left( v_j(g) - \mu w_j^* - c d(j) \right)}
			+ \sum\limits_{j \in \mathbb{S}_+(g^*) \setminusx \{1\}}{\left( - \delta \mu + c \ceil{\delta \mu / c} \right)}\\
			&= \sum\limits_{j \in \mathbb{S}_+(g^*)}{(v_j(g^*) - \mu w_j^* - cd(j))}
			+ (\Delta - K^* \mu - \delta)\delta
			- \delta (\Delta - \mu (K_{max}^* + 1) - \delta)\\
			&\phantom{xxx}
			- c \floor[\Big]{\frac{\delta \mu}{c}(K_{max}^* - K^* + 1)}
			+ (K^* - 1) (c\ceil{\delta \mu / c} - \delta \mu)\\
			&= \E[RetailerWelfare]
			+ (\Delta - K^* \mu - \delta)\delta
			- \delta (\Delta - \mu (K_{max}^* + 1) - \delta)\\
			&\phantom{xxx}
			- c \floor[\Big]{\frac{\delta \mu}{c}(K_{max}^* - K^* + 1)}
			+ (K^* - 1) (c\ceil{\delta \mu / c} - \delta \mu)\\
			&\approx \E[RetailerWelfare],
	\end{align*}
	where the last approximation is obtained by assuming divisibility in floor/ceil operators
	in the obtained expressions.
	
	Having collected the changes to each of the welfare components, we can establish that
	the total expected welfare changes in response to the first supplier's increasing its mean supply
	by $\delta$ as
	\begin{align*}
		\E[Welfare]' \approx \E[Welfare] + \delta(\Delta - \mu (K_{max}^* - K + 1) - \delta / 2).
	\end{align*}
	Inside the second factor in the obtained expression, $\Delta$ is the dominating term (as the
	chain is large), so, in general, the change to the total welfare is positive.
	\qed
\endproof

Notice that the expression for social welfare in Theorem~\ref{thm:social-welfare-varying-mu}---unlike
the analogous expression in Theorem~\ref{thm:nash-equilibrium-efficiency} for the case of identical
suppliers---depends on which equilibrium retailers arrive at and, more specifically, on the number of
suppliers being active at that equilibrium.

\section{Conclusion}

In this work, we have considered strategic formation of supply chains with strategic suppliers---who set prices anticipating retailer response---and strategic retailers---who link to suppliers maximizing expected payoffs and being driven by both supply uncertainty and the set prices. Our major findings are that (i) formed supply chain equilibria are inefficient w.r.t. centrally planned supply chains, and (ii) different ways to improve supply uncertainty have different effects upon welfare---increasing mean supply is universally good, while decreasing supply variance lowers consumer surplus.

\ACKNOWLEDGMENT{%
The work is supported in part by the Rockefeller Foundation under grant~{2017 PRE 301}.%
}

\bibliographystyle{informs2014}
\bibliography{main-alias,main}

\begin{APPENDICES}

\section{Proofs}

\proof{\hypertarget{thm:expected-retailer-utility-proof}{Proof of Lemma~\ref{thm:expected-retailer-utility}}:}
	\begin{align*}
		\E[&u_i(g, w)]\\
		&= \E\left[ \sum\limits_{j \in N(i)}{
			\left( \Big( \Delta - \sum\limits_{k \in \mathbb{S}_+(g)}{S_k} - w_j \Big) \frac{S_j}{d(j)} - c \right)
		} \right]
		= \sum\limits_{j \in N(i)}{
			\Bigg( (\Delta - w_j) \frac{\E[S_j]}{d(j)} - \frac{\E\Big[S_j \sum\limits_{k \in \mathbb{S}_+(g)}{S_k} \Big]}{d(j)} - c \Bigg)
		}\\[0.12in]
		&= (\text{as } S_j \text{ are i.i.d.}, j \in \mathbb{S}_+(g), \text{ and } \Var[S_j] = \E[S_j^2] - \E^2[S_j])\\[0.14in]
		&= \sum\limits_{j \in N(i)}{
			\left( \frac{\mu (\Delta - w_j)}{d(j)} - \frac{\sum\limits_{k \in \mathbb{S}_+(g)}{\E[S_j] \E[S_k]} + \Var[S_j]}{d(j)} - c \right)
		}
		= \sum\limits_{j \in N(i)}{
			\left( \frac{\mu (\Delta - w_j)}{d(j)} - \frac{\mu^2 K + \sigma^2}{d(j)} - c \right)
		}\\[0.1in]
		&= \sum\limits_{j \in N(i)}{
			\left( \frac{\mu (\Delta - \mu K) - \sigma^2 - \mu w_j)}{d(j)} - c \right)
		}
		= \sum\limits_{j \in N(i)}{
			\left( \frac{v(K) - \mu w_j}{d(j)} - c \right)
		},
	\end{align*}
	where $K = K(g) = |\mathbb{S}_+(g)|$ is the number of active suppliers in $g$.
	\qed
\endproof

\vspace{0.1in}

\proof{\hypertarget{thm:network-formation-nash-equilibrium-char-proof}{Proof of Theorem~\ref{thm:network-formation-nash-equilibrium-char}}:}
	From Theorem~\ref{thm:active-sups-at-equilibrium}, we know that the $K$ cheapest suppliers
	are active at an equilibrium. For active supplier $j$ from that supplier set, the marginal benefit
	of linking to it (by a vacant retailer whose number is sufficiently large) should be non-negative
	$$
		\frac{v(K) - \mu w_j}{d(j)} - c \geq 0;
	$$
	and the marginal benefit of creating an extra link to it (by any retailer) should be negative
	$$
		\frac{v(K) - \mu w_j}{d(j) + 1} - c < 0.
	$$
	Combining the two obtained inequalities, we get
	$$
		d(j) = \floor[\Big]{ \frac{v(K) - \mu w_j}{c} }.
	$$
	\qed
\endproof

\section{Key Metrics of Formed Supply Chains}

\newcommand{\uberwidth}[0]{0.25\linewidth}
\newcommand{\ubergap}[0]{\hspace{0.1in}}
\begin{figure}[h]
	\centering
	\subfloat[][$K^*$ vs. $\mu$]{\includegraphics[width=\uberwidth]{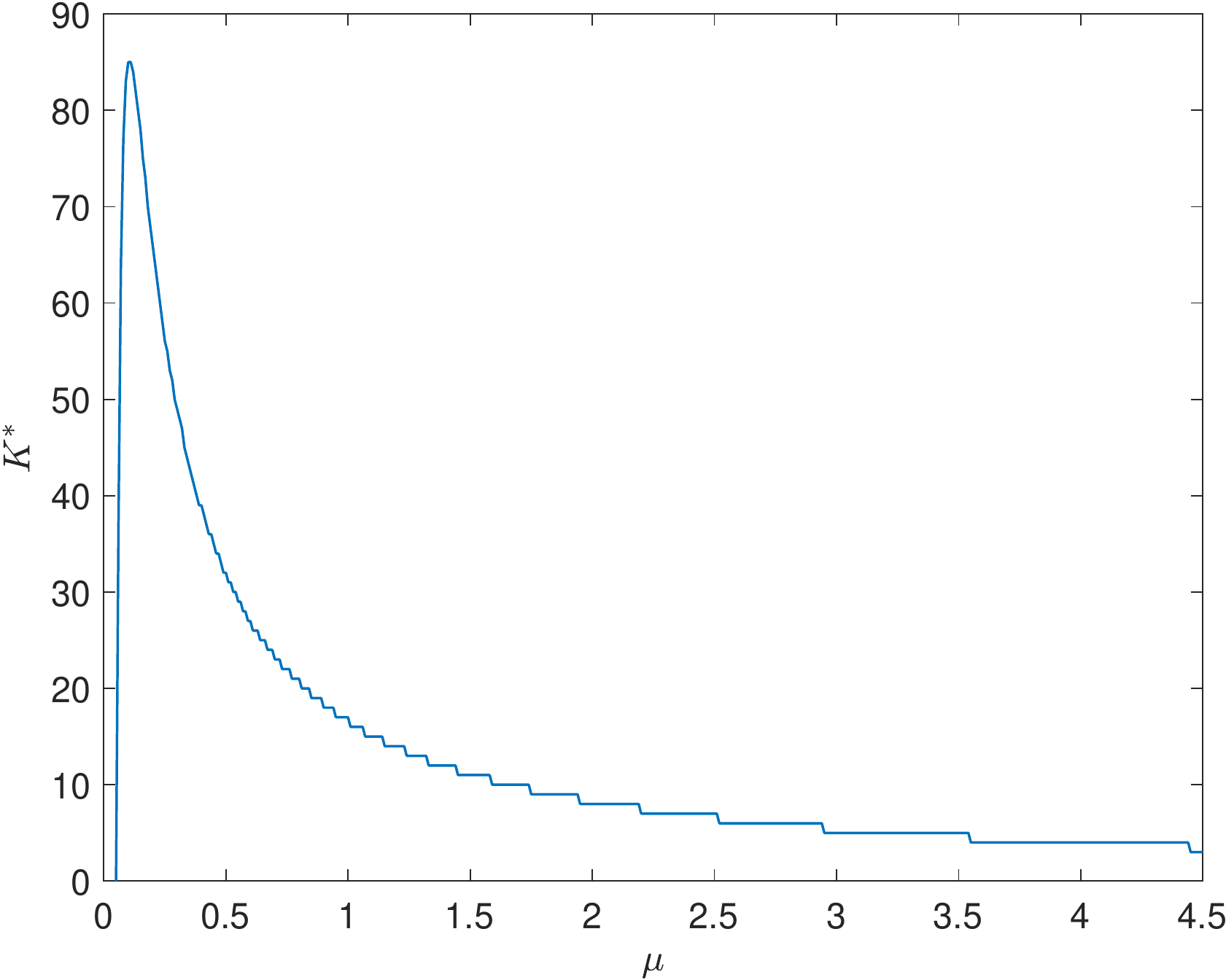}} \ubergap
	\subfloat[][$d(j)$ vs. $\mu$]{\includegraphics[width=\uberwidth]{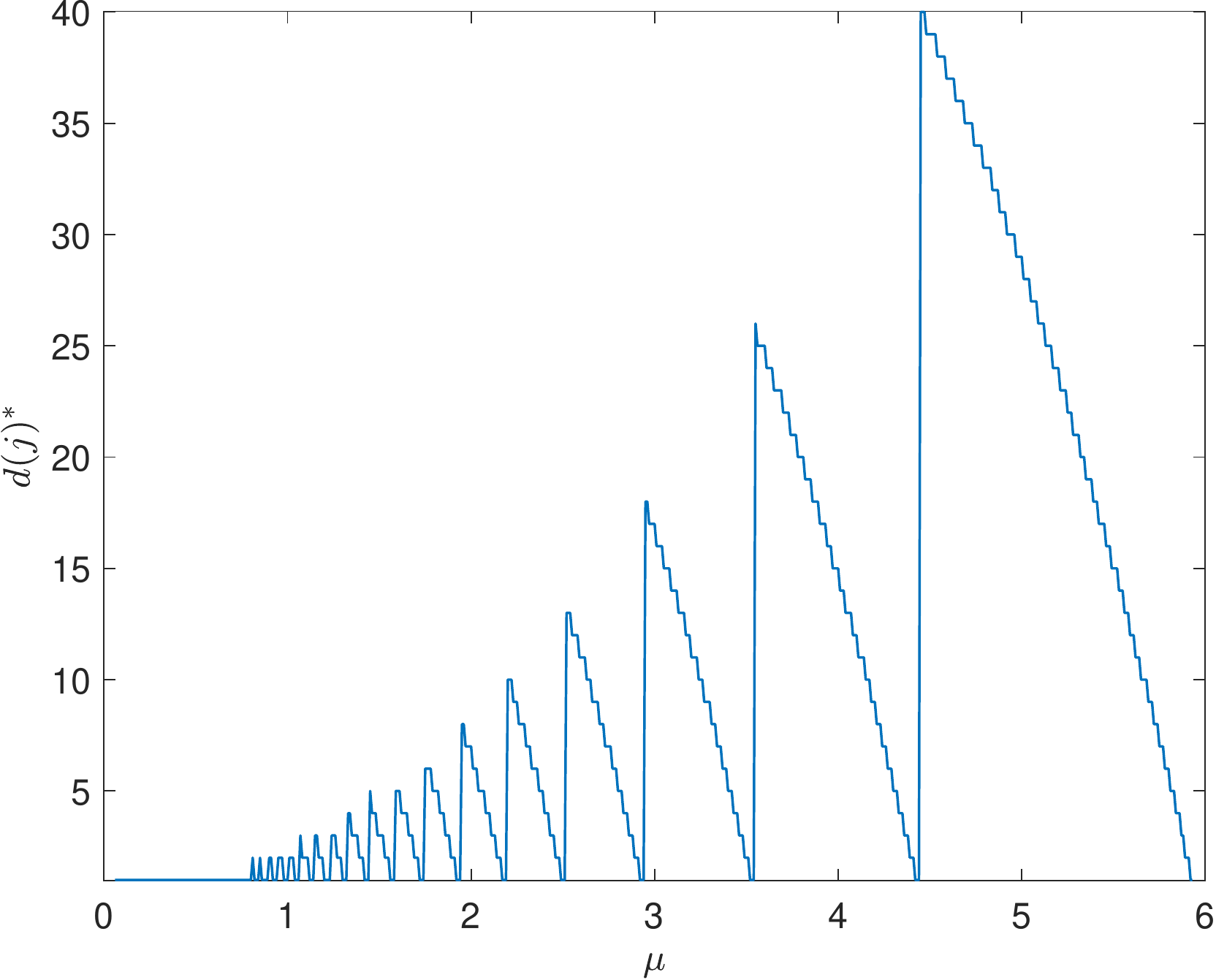}} \ubergap
	\subfloat[][$d(j)$ vs. $\sigma^2$]{\includegraphics[width=\uberwidth]{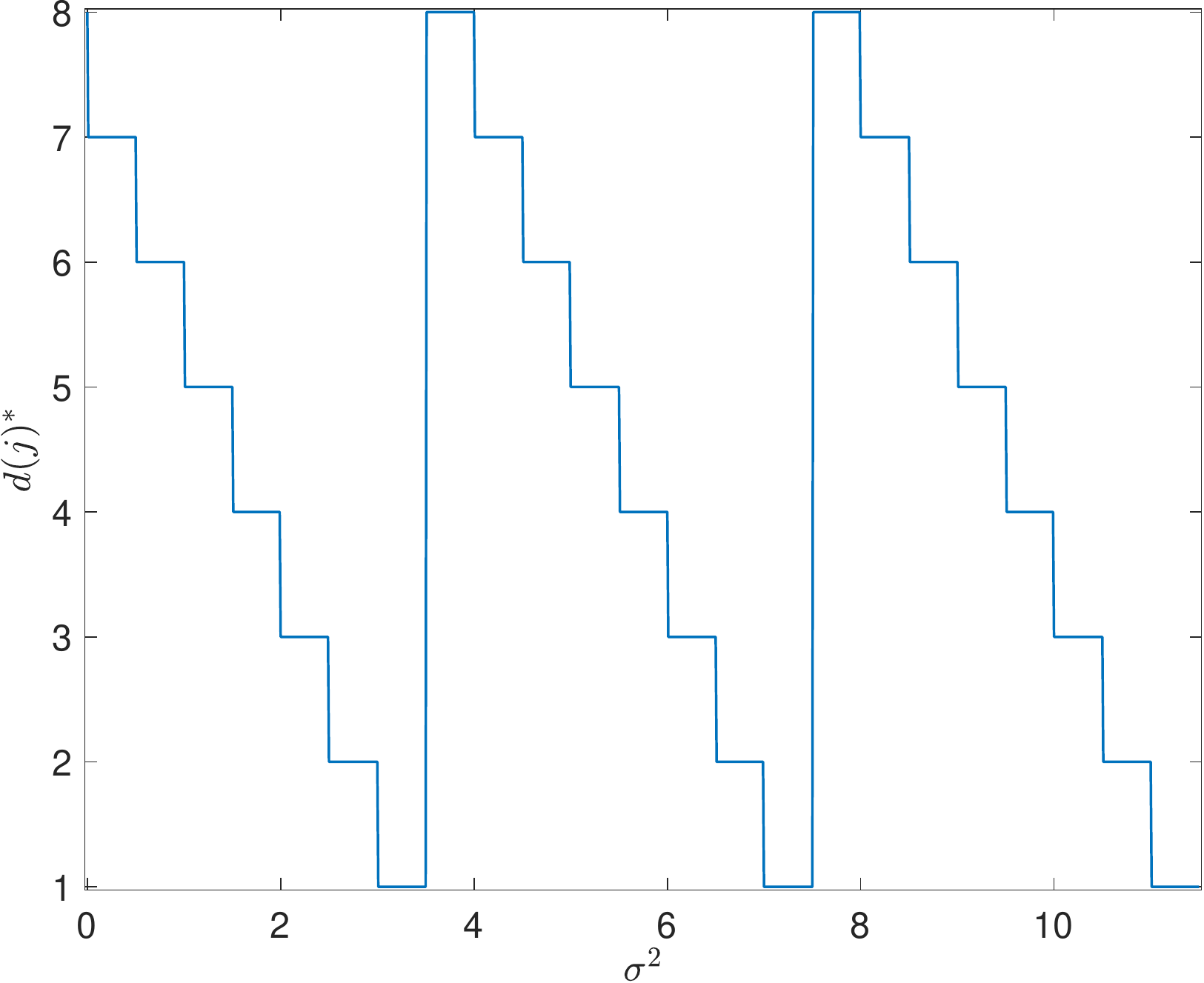}}
     \\
	\subfloat[][$K^{opt}$ vs. $\mu$]{\includegraphics[width=\uberwidth]{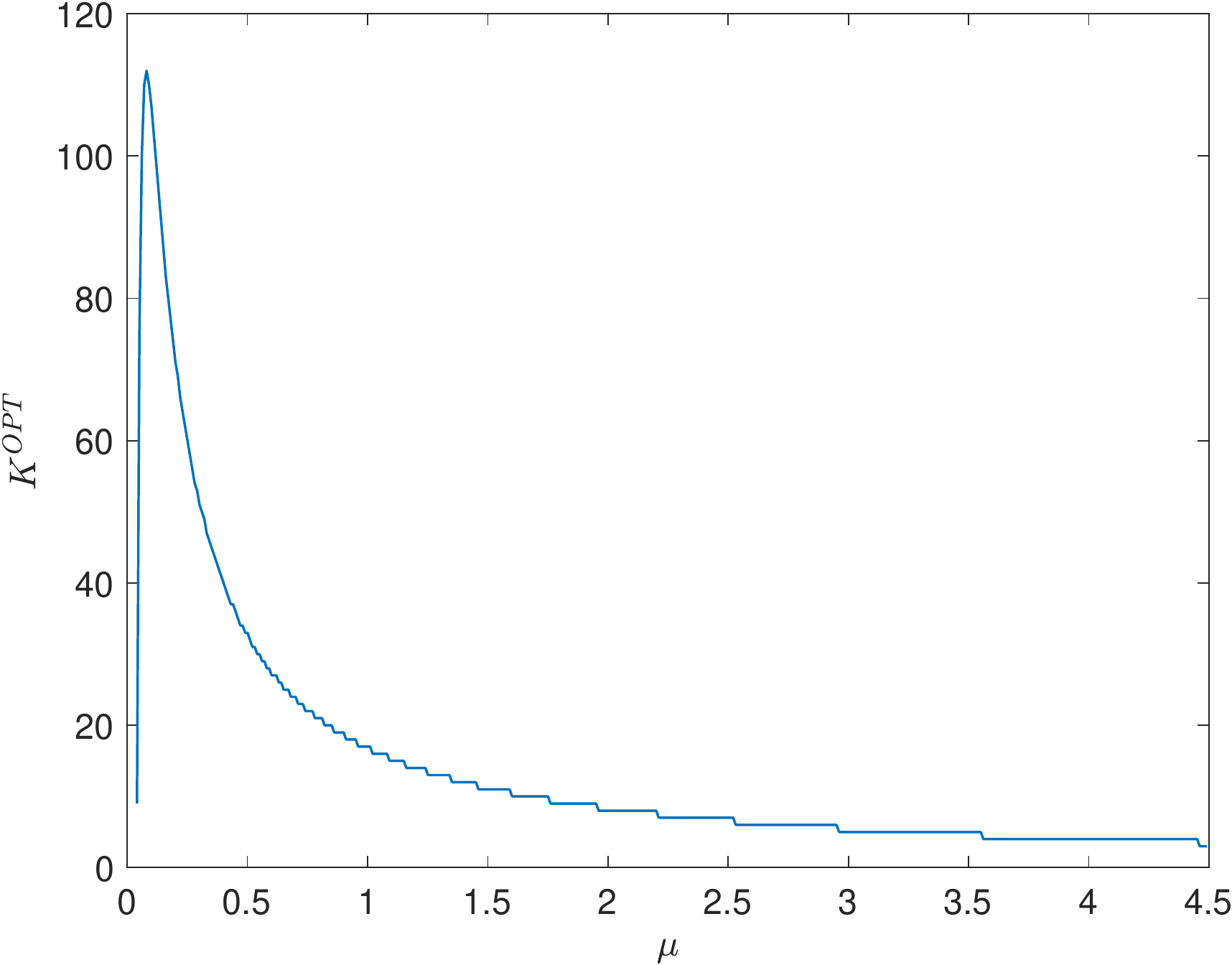}} \ubergap
	\subfloat[][$\E~Welfare^*$ vs. $\mu$]{\includegraphics[width=\uberwidth]{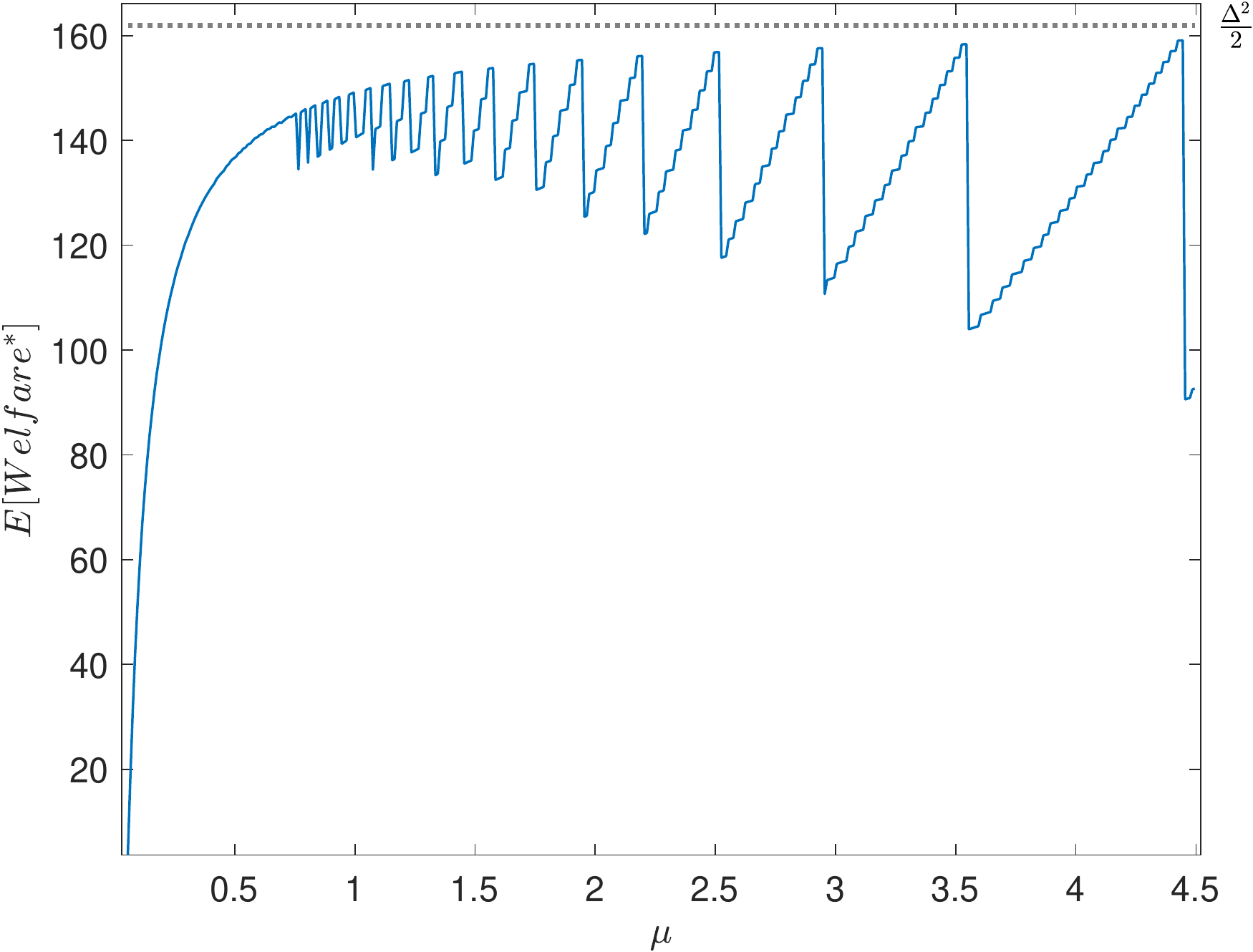}} \ubergap
	\subfloat[][$\E~Welfare^*$ vs. $\sigma^2$]{\includegraphics[width=\uberwidth]{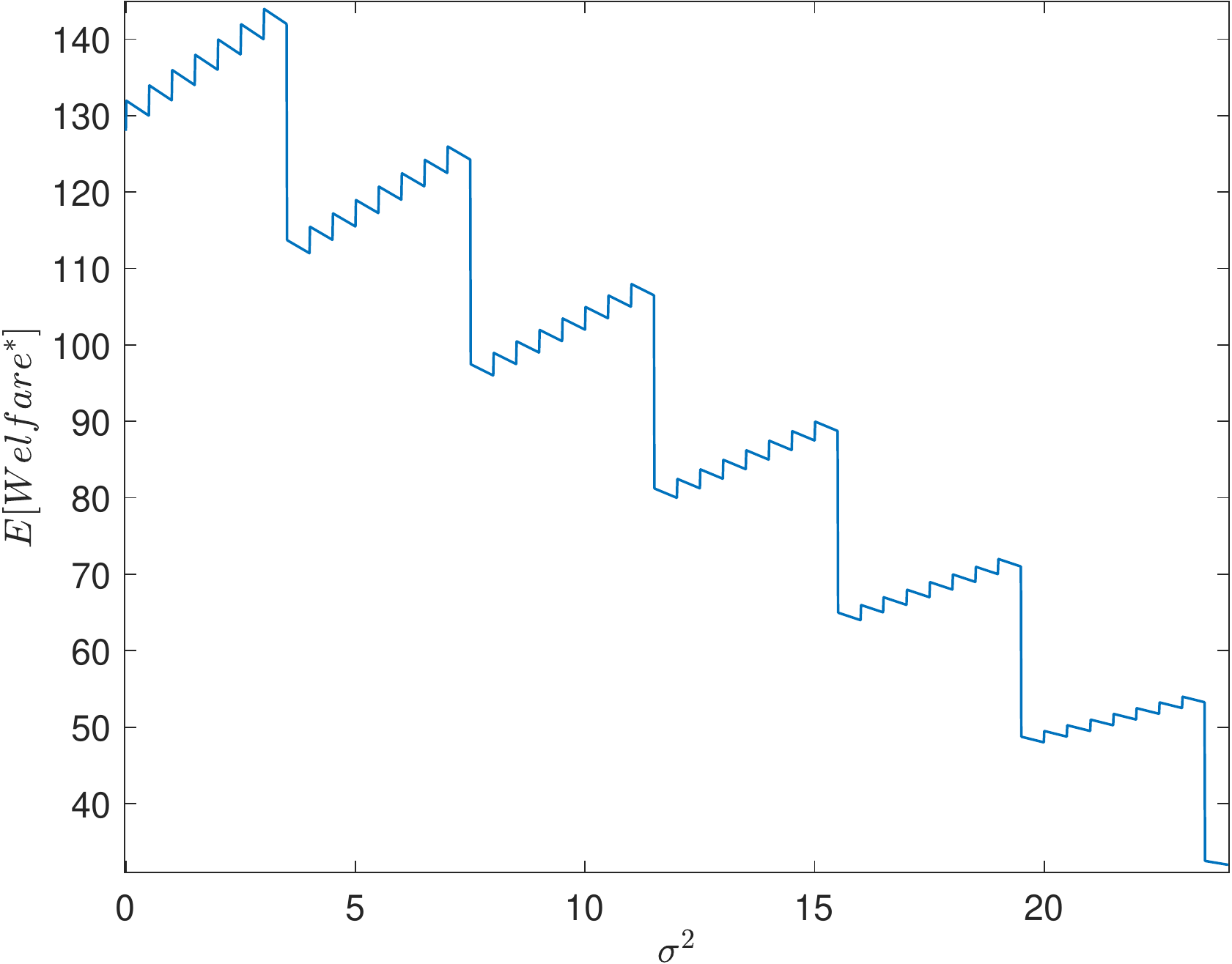}}
     \\
	\subfloat[][$\E~Welfare^{opt}$ vs. $\mu$]{\includegraphics[width=\uberwidth]{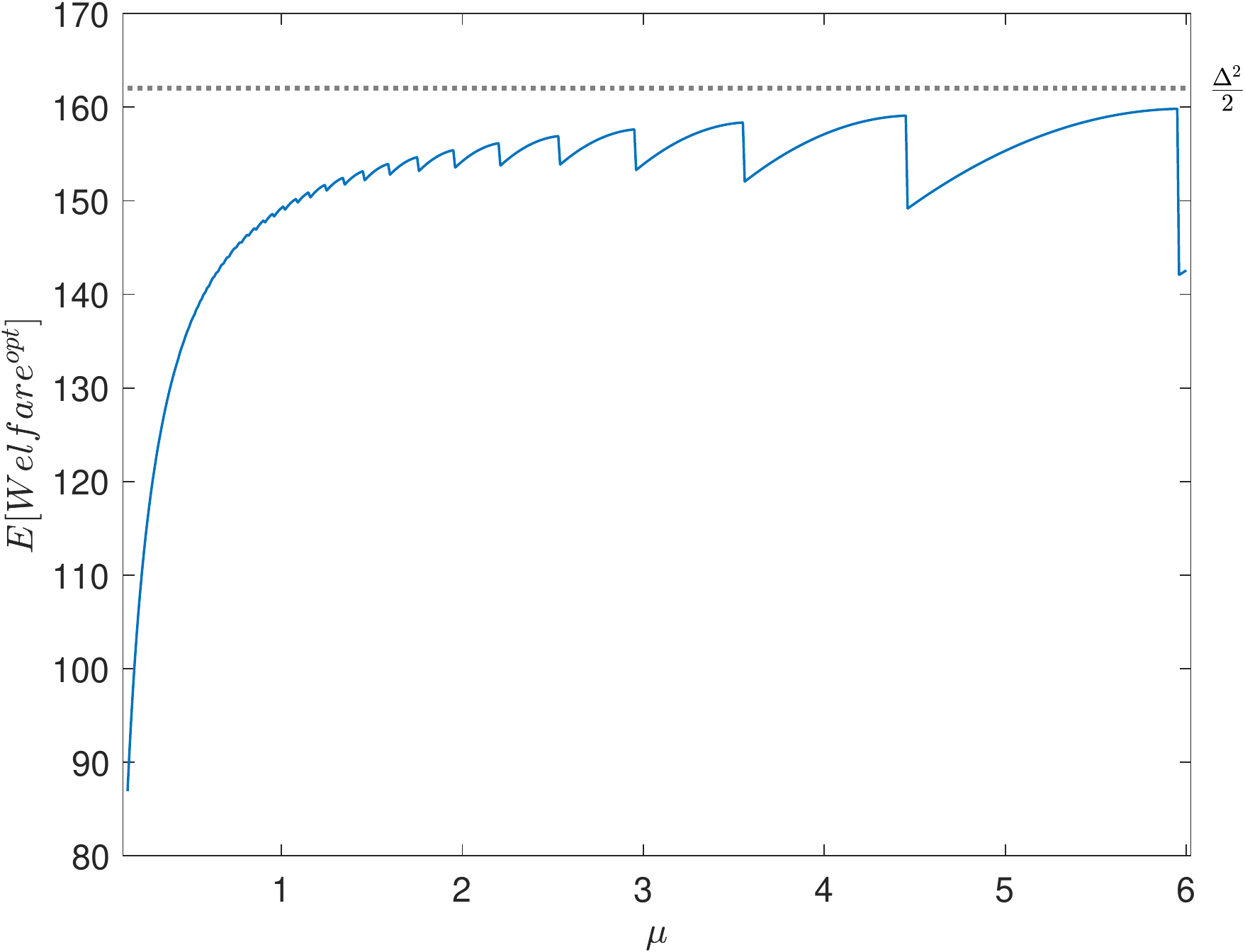}} \ubergap
	\subfloat[][$PoS$ vs. $\mu$]{\includegraphics[width=\uberwidth]{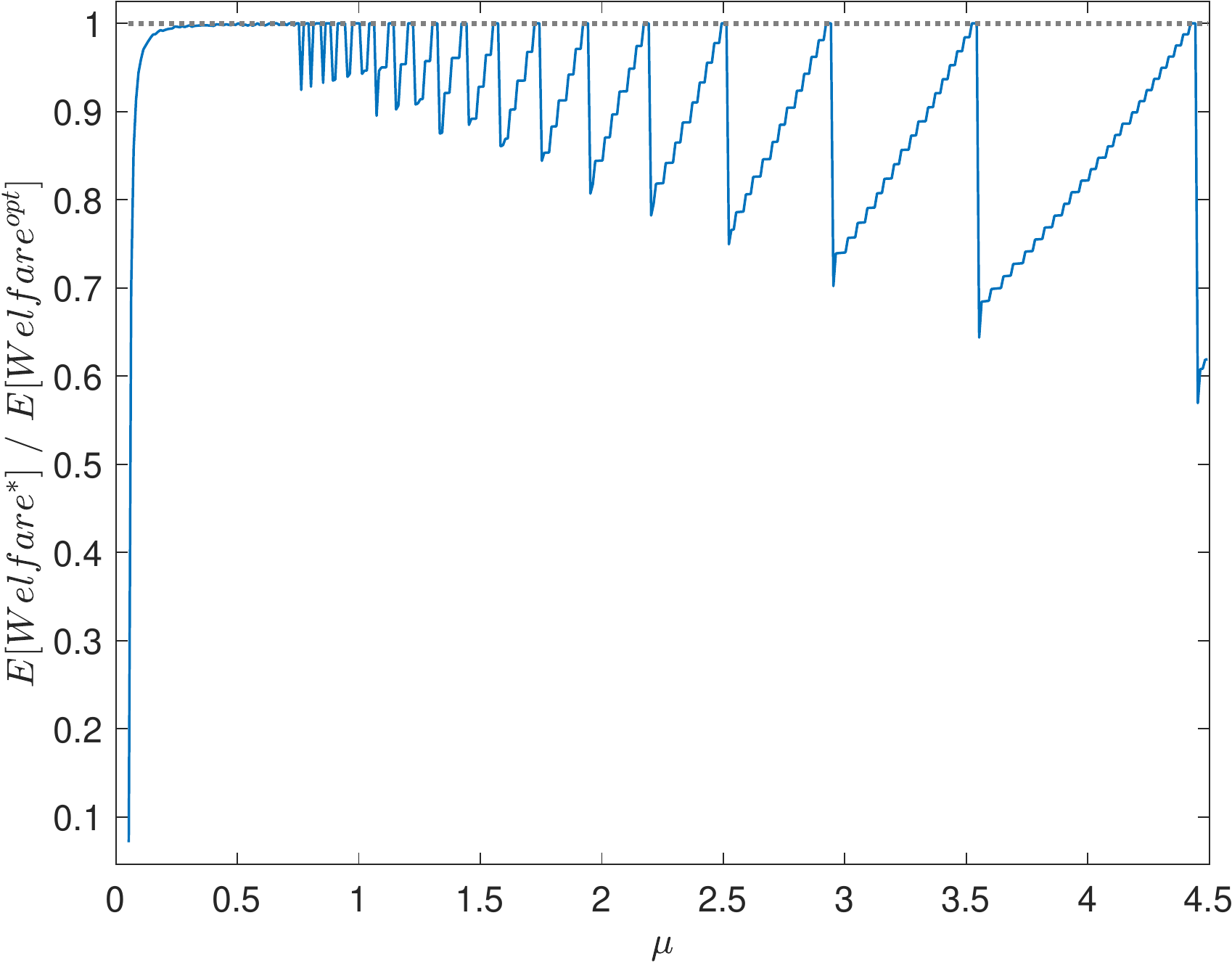}} \ubergap
	\subfloat[][$PoS$ vs. $\sigma^2$]{\includegraphics[width=\uberwidth]{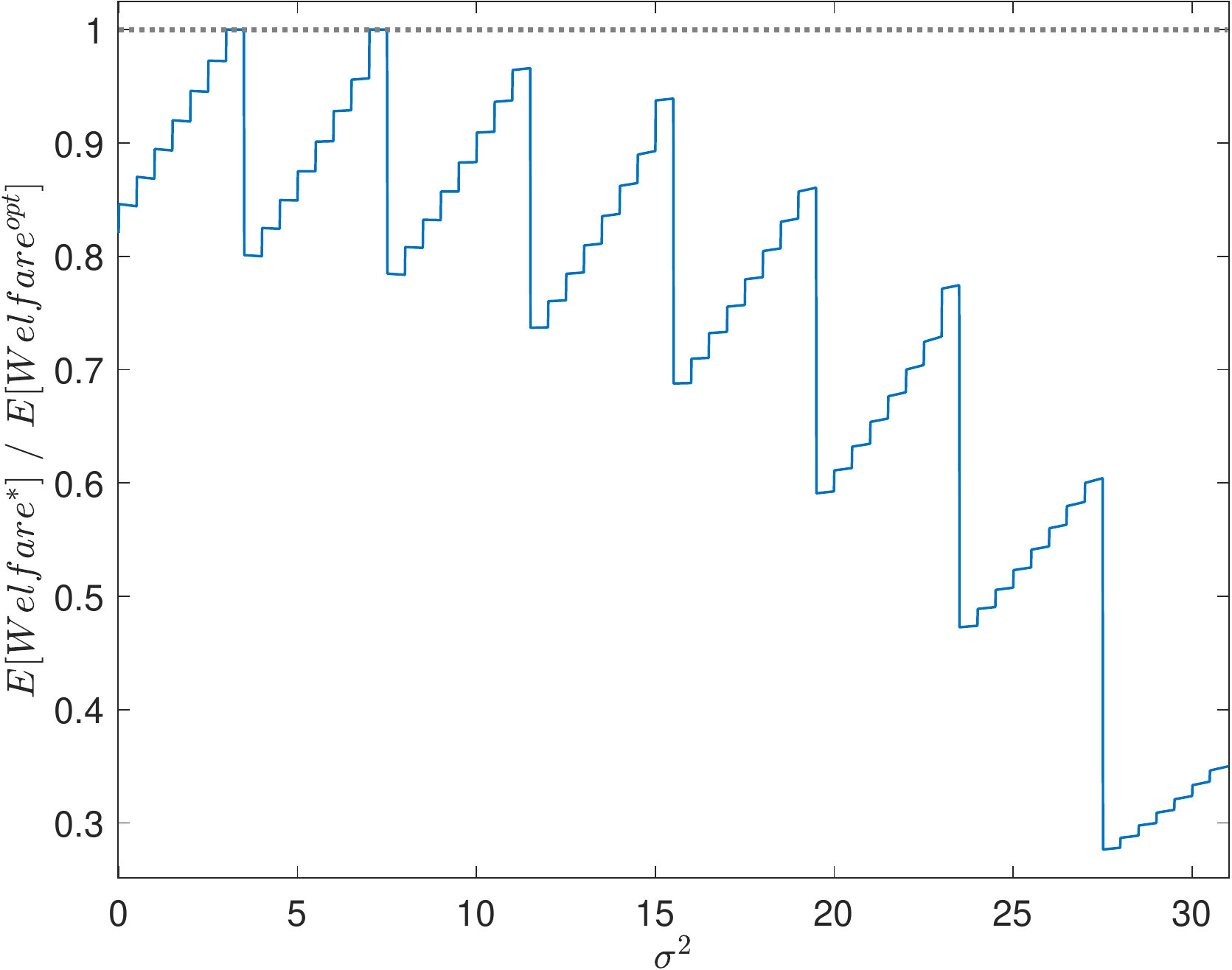}}\\
	\subfloat[][$PoS$ vs. $m$]{\includegraphics[width=\uberwidth]{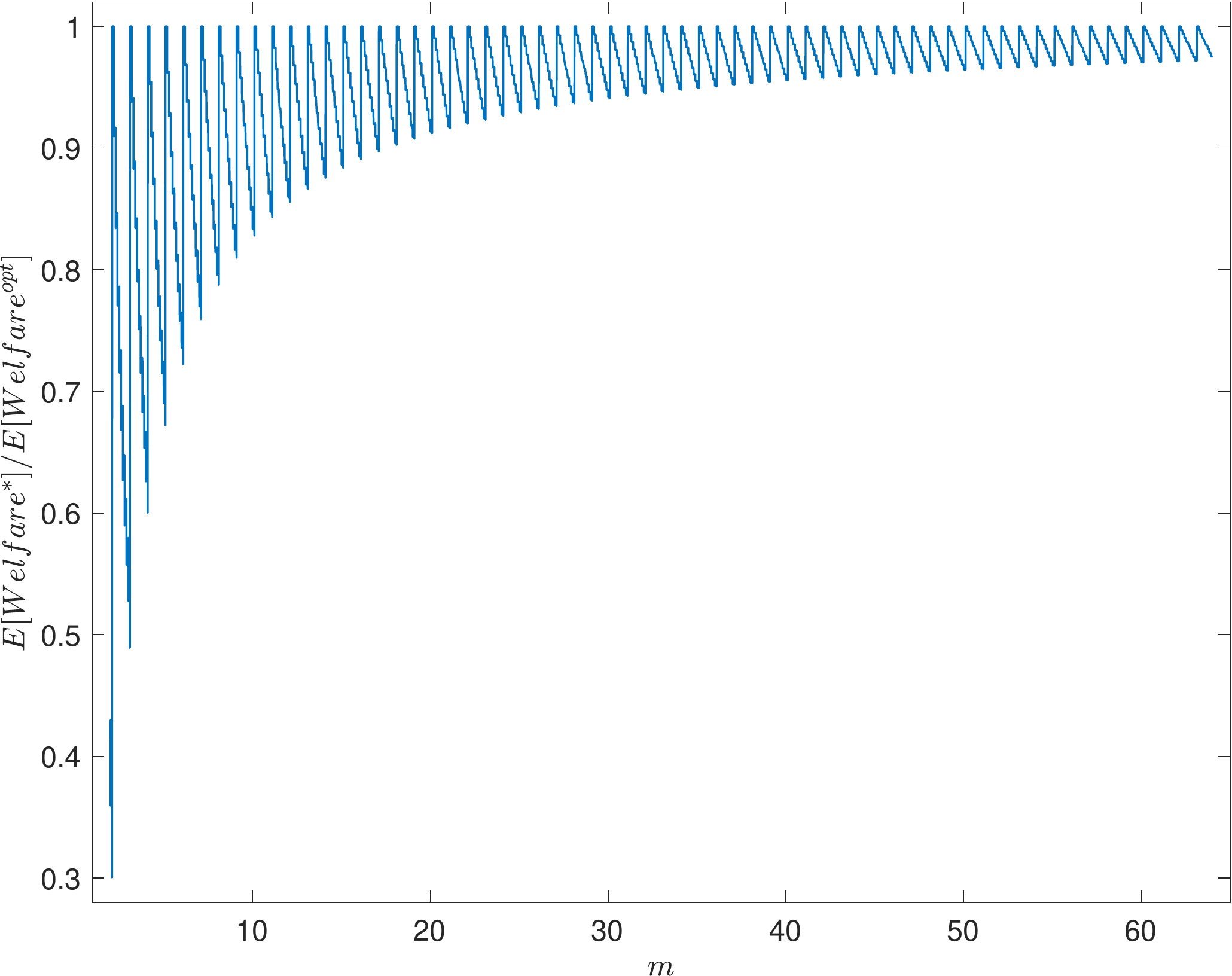}}
	\caption{
		Dependence on supplier degree $d(j)$, number of active suppliers $K$,
		expected social welfare $\E[Welfare]$, and the price of stability (PoS)
		$\E[Welfare^{*}]~/~\E[Welfare^{opt}]$
		at an equilibrium ($^*$) and for the central planner ($^{opt}$) on
		supply mean $\mu$, variance $\sigma^2$, and the number of suppliers $m$.
	}
\end{figure}

\end{APPENDICES}

\end{document}